\documentclass[aps,prb,twocolumn,superscriptaddress,nobibnotes]{revtex4-2}

\usepackage{lipsum}%lorem ipsum
\usepackage{xcolor}%coloured text
\usepackage{graphicx}% Include figure files
\usepackage{dcolumn}% Align table columns on decimal point
\usepackage{bm}% bold math
\usepackage{amsfonts}
\usepackage{amsmath,amssymb}
\usepackage{siunitx}
\usepackage{dsfont}%indicator function font
\usepackage{placeins}
\usepackage{braket}
\usepackage{xcolor}
\usepackage[normalem]{ulem}

\usepackage{tikz}

\definecolor{bluepoli}{RGB}{0,36,179}
\definecolor{redpoli}{RGB}{204,0,51}
\definecolor{greenpoli}{RGB}{45,137,0}
\definecolor{purplepoli}{RGB}{153,102,204}
\definecolor{azzurropoli}{RGB}{51,53,204}
\definecolor{orangepoli}{RGB}{255,124,17}
\usepackage{hyperref}
\hypersetup{
	colorlinks,
	linkcolor={bluepoli}, %blue poli
	citecolor={redpoli}, %rosso
	urlcolor={redpoli} % 
}

\usepackage[capitalise]{cleveref}
\begin{document}

%\title{Machine-learned tuning of artificial Kitaev chains to Majorana sweet spots}
\title{Machine-learned tuning of artificial Kitaev chains\\ from tunneling-spectroscopy measurements}

\author{Jacob Benestad}
\email{jacob.d.benestad@ntnu.no}
\affiliation{Center for Quantum Spintronics, Department of Physics, Norwegian University of Science and Technology, 7491 Trondheim, Norway}

\author{Athanasios Tsintzis}
\affiliation{Division of Solid State Physics and NanoLund, Lund University, S-22100 Lund, Sweden}

\author{Rubén Seoane Souto}
\affiliation{Instituto de Ciencia de Materiales de Madrid, Consejo Superior de Investigaciones Científicas (ICMM-CSIC), Madrid, Spain}
\affiliation{Division of Solid State Physics and NanoLund, Lund University, S-22100 Lund, Sweden}

\author{Martin Leijnse}
\affiliation{Division of Solid State Physics and NanoLund, Lund University, S-22100 Lund, Sweden}

\author{Evert van Nieuwenburg}
\affiliation{Lorentz Institute and Leiden Institute of Advanced Computer Science,
	Leiden University, P.O. Box 9506, 2300 RA Leiden, The Netherlands}

\author{Jeroen Danon}%
\affiliation{Center for Quantum Spintronics, Department of Physics, Norwegian University of Science and Technology, 7491 Trondheim, Norway}

\date{\today}

\begin{abstract}
We demonstrate reliable machine-learned tuning of quantum-dot-based artificial Kitaev chains to Majorana sweet spots, using the covariance matrix adaptation algorithm.
We show that a loss function based on local tunnelling-spectroscopy features of a chain with two additional sensor dots added at its ends provides a reliable metric to navigate parameter space and find points where crossed Andreev reflection and elastic cotunneling between neighbouring sites balance in such a way to yield near-zero-energy modes with very high Majorana quality.
We simulate tuning of two- and three-site Kitaev chains, where the loss function is found from calculating the low-energy spectrum of a model Hamiltonian that includes Coulomb interactions and finite Zeeman splitting.
In both cases, the algorithm consistently converges towards high-quality sweet spots.
Since tunnelling spectroscopy provides one global metric for tuning all on-site potentials simultaneously, this presents a promising way towards tuning longer Kitaev chains, which are required for achieving topological protection of the Majorana modes.
\end{abstract}

\maketitle

%%%%%%%%%%%%%%%%
% INTRODUCTION %
%%%%%%%%%%%%%%%%

\section{Introduction}
\label{sec:introduction}

Realization and control of Majorana bound states (MBSs), which remain much sought after due to their non-local and topological nature, are a topic of intense study within condensed matter physics \cite{beenakker_search_2013, beenakker_search_2020, aasen_milestones_2016, alicea_new_2012, alicea_non-abelian_2011, nayak_non-abelian_2008}. The simple Kitaev model gives one example of a topological material where MBSs can be found \cite{kitaev_unpaired_2001}, and recreating the ingredients of the Kitaev model with engineered hybrid devices is a very promising avenue in this pursuit \cite{flensberg_engineered_2021}. While a lot of work has focused on making artificial chains using semiconductor nanowires proximitized by superconductors \cite{deng_majorana_2016, deng_anomalous_2012, finck_anomalous_2013, lutchyn_majorana_2010, mourik_signatures_2012, nichele_scaling_2017, oreg_helical_2010, prada_andreev_2020}, device imperfections, such as material disorder, seem currently to be a major challenge for further progress in this direction \cite{liu_zero-bias_2012, pikulinZerovoltageConductancePeak2012, kellsNearzeroenergyEndStates2012,  zhang_ballistic_2017, reegZeroenergyAndreevBound2018, woodsZeroenergypinning2019, pan_physical_2020, ahn_estimating_2021, das_sarma_disorder-induced_2021, yu_non-majorana_2021, hessLocalNonlocalQuantum2021a, cayaoConfinementinducedZerobiasPeaks2021, hessTrivialAndreevband2023}.

An alternative is to rather use arrays of alternating normal and proximitized quantum dots to build an artificial Kitaev chain, which could circumvent some of the material disorder issues \cite{sau_realizing_2012, leijnse_parity_2012, miles_kitaev_2023, tsintzis_creating_2022, tsintzis_roadmap_2023, luna_flux-tunable_2024, liu_fusion_2023, souto_subgap_2024}.
Signatures of so-called ``Poor Man's Majoranas'' (PMMs), which are non-local but lack topological protection, have been demonstrated in a two-site minimal artificial Kitaev chain \cite{dvir_realization_2023, haaf_engineering_2023, zatelli_robust_2023}.
An important challenge faced in this field is to make and tune devices that are long enough that they can host topologically protected Majoranas \cite{sau_realizing_2012}. 
Recent progress along these lines includes control over three-site devices \cite{bordin_crossed_2024} and the observation of signatures of tuning to a Majorana sweet spot in three-site devices~\cite{bordin_signatures_2024}.

To tune these artificial Kitaev chains to a sweet spot, where an even--odd ground-state degeneracy coincides with maximal Majorana quality (which we will define later) of the part of the corresponding excitation that lives on the outer sites of the chain, it is necessary to tune the system parameters such that crossed Andreev reflection (CAR) and elastic cotunneling (ECT) processes via the proximitized dots are pairwise balanced in each link in the chain \cite{liu_tunable_2022, liu_enhancing_2023}.
In experiment, one faces the challenge of imperfect information about said tuning parameters, due to microscopic details such as material imperfections and varying gate lever arms. In practice, tuning has so far been done by working with pairs of sites at a time, since there are too many variables for navigating a global metric manually, where all dots are tuned simultaneously. Such sequential manual tuning across the chain can be time-consuming, and at the same time one must be careful that the overall tuning remains correct each time a new pair is tuned \cite{bordin_crossed_2024, bordin_signatures_2024}. Pairwise tuning becomes especially precarious for the case of strong tunnel couplings, where multi-dot renormalization effects can be expected. 
With single CAR and ECT processes thus depending on parameters all along the chain, it will be difficult to tune the system while only considering pairs of dots. Thus, tuning all dots simultaneously would be advantageous, and the fact that this would be very difficult for a human to do motivates us to investigate machine-learning methods to perform automated tuning. 

\begin{figure*}[t]
    \centering
    \includegraphics[width=\linewidth]{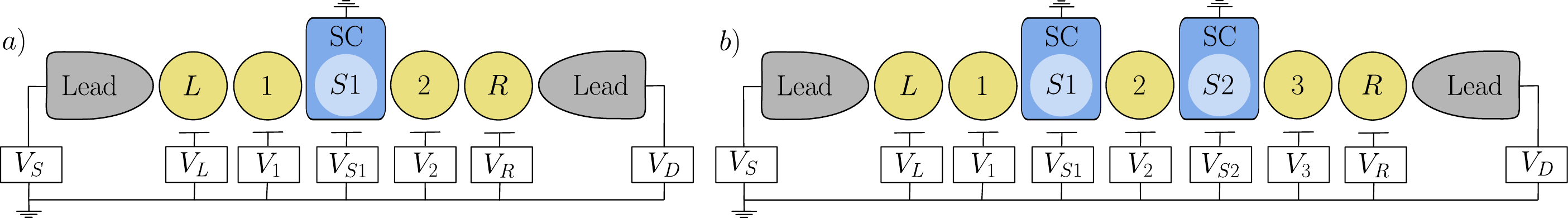}
    \caption{Sketch of the artificial Kitaev chain systems considered in this paper. In addition to the sensor dots marked $L/R$ on either side of the Kitaev chain, the array consists of (a) $3$ dots or (b) $5$ dots, where every other dot is proximitized by a superconductor (blue square) to create an Andreev bound state that can mediate CAR and ECT processes between the normal dots.}
    \label{fig:setup}
\end{figure*}

Such automatic tuning has already been proposed as a way to mitigate disorder in bulk hybrid nanowire sytems, by using an Aharonov--Bohm interferometer to get a metric of the Majorana quality that can be used as a global loss function \cite{thamm_machine_2023}.
A simpler and more practical probe of Majorana quality could be to perform a tunneling spectroscopy experiment while adding an extra sensor dot at the end of the system \cite{deng_majorana_2016, prada_measuring_2017, clarke_experimentally_2017}, which was recently investigated in the context of artifical Kitaev chains based on quantum dot arrays \cite{souto_probing_2023}. Work has also been done on the machine-learned tuning of dot-based artificial Kitaev chains, in the context of using generative neural networks to automatically classify avoided crossings in charge stability diagrams \cite{koch_adversarial_2023}.

In this paper, we demonstrate automatic tuning in simulated two- and three-site artificial Kitaev chains using the covariance matrix adaptation evolutionary strategy (CMA-ES) \cite{hansen_cma_2023, hansen_completely_2001, hansen_reducing_2003} with tunneling spectroscopy through sensor dots at both ends of the chain as a metric to navigate parameter space.
We first apply the tuning algorithm to a minimal (two-site) Kitaev chain, where locations of the sweet spots can be relatively easily determined by numerical calculations for a given set of tight-binding parameters~\cite{tsintzis_creating_2022}, and we show how the algorithm indeed converges towards these points.
We then consider a three-site chain, where the locations of the sweet spots are not known a priori.
Also in this case, the algorithm converges to points with vanishing even--odd ground-state splitting and very high Majorana quality.
We thus demonstrate the capability of the CMA-ES algorithm to navigate a large tuning parameter space and efficiently locate Majorana sweet spots, based on simple tunneling-spectroscopy data.

The rest of this paper is structured as follows:
In \cref{sec:model} we introduce the Hamiltonian for our quantum-dot-based artificial Kitaev model, and we explain the basic working of the CMA-ES algorithm used for the automatic tuning.
In \cref{sec:results} we present our results for the automatic tuning, first for a two-site model in \cref{sec:twosite}, and a three-site model in \cref{sec:threesite}.
Finally, we present a short conclusion in \cref{sec:conclusion}.

% END INTRODUCTION %

%%%%%%%%%
% MODEL %
%%%%%%%%%
%\FloatBarrier
\section{Model}
\label{sec:model}

We consider a similar setup to that presented in Refs.~\cite{souto_probing_2023, tsintzis_roadmap_2023}: a linear array of quantum dots where every other dot is proximitized by a superconductor, and with the addition of extra sensor dots at each end of the chain.
To make the optimization more robust, we probe the low-energy physics of the system first with the sensor dot added to the left side of the array and then a second time with the sensor dot on the right side.
In an experiment, such freedom of changing sides for the sensor dot could be achieved by defining extra dots on both ends of the Kitaev chain and then ``removing'' the one that is not being used by lowering its outer barrier such that the dot becomes incorporated as a part of the lead.

We describe the full array of dots with the Hamiltonian
\begin{equation}
\label{eq:hamfull}
    H^{(n,s)} = H_{\text{Kit}}^{(n)} + H_\text{S}^{(n,s)},
\end{equation}
where $s=L,R$ indicates on which side the sensor dot is located and $n$ labels the number of ``Kitaev-chain'' sites, i.e., the number of normal (non-proximitized) quantum dots.

The first term in \cref{eq:hamfull} describes the Kitaev-chain part of the system, which includes normal dots and dots proximitized by a superconductor, together with spin-conserving and spin-non-conserving hopping between them,
\begin{align}
    H_{\text{Kit}}^{(n)} &= \sum_{j=1}^n \sum_\sigma \varepsilon_{j \sigma} d_{j\sigma}^\dagger d_{j \sigma} + \sum_{j=1}^n \frac{U}{2} n_j (n_j-1) \label{eq:hkit}\\
    & + \sum_{j=1}^{n-1} \Big[ \varepsilon_{Sj} n_{Sj} + \Delta \big( d_{Sj\uparrow}^\dagger d_{Sj \downarrow}^\dagger +  d_{Sj\downarrow} d_{Sj \uparrow}\big) \Big] \nonumber\\
    & + \sum_{j=1}^{n-1} \sum_\sigma \Big[ t_{j, Sj} \,  d_{j\sigma}^\dagger d_{Sj \sigma} + t_{Sj ,j+1} \, d_{Sj \sigma}^\dagger d_{(j+1)  \sigma} \nonumber\\
    & +  \beta_\sigma \big( t_{j, Sj}^{\text{SO}} \,  d_{j\sigma}^\dagger d_{Sj \bar{\sigma}} + t_{Sj , j+1}^{\text{SO}} \, d_{Sj\sigma}^\dagger d_{(j+1) \bar{\sigma}} \big) + \text{H.c.} \Big],\nonumber
\end{align}
where $d_{j\sigma}^\dagger$ ($d_{Sj\sigma}^\dagger$) is the creation operator for an electron with spin $\sigma=\uparrow,\downarrow$ on (proximitized) dot $j$ ($Sj$), and $n_j=d_{j\uparrow}^\dagger d_{j \uparrow}+d_{j\downarrow}^\dagger d_{j \downarrow}$ ($n_{Sj}=d_{Sj\uparrow}^\dagger d_{Sj \uparrow}+d_{Sj\downarrow}^\dagger d_{Sj \downarrow}$) is the number operator on that dot.
The normal dots are described by the first line in \cref{eq:hkit}.
Their on-site single-particle energy $\varepsilon_{j\sigma}=\varepsilon_j + E_{Z}\, \delta_{\sigma,\downarrow}$ is spin-dependent, where $E_Z$ is the Zeeman energy and $\varepsilon_j$ can be tuned by nearby electrostatic gates $V_j$, and the charging energy associated with a doubly occupied normal dot is given by $U$.
The second line in \cref{eq:hkit} describes the proximitized dots, labeled $Sj$.
The proximity of the superconductor is expected to result in a strong reduction of the effective on-site $g$-factors, which we incorporate by using spin-independent potentials $\varepsilon_{Sj}$, for simplicity.
The superconductor will also efficiently screen the on-site Coulomb interactions on these dots, which we thus neglect here~\footnote{Including small Coulomb interactions on the proximitized dots does not change the qualitative results~\cite{tsintzis_creating_2022}}.
The proximity-induced pairing potential $\Delta$ is assumed to be real and equal for all proximitized dots.
Finally, the last two lines in \cref{eq:hkit} describe the coupling between the dots.
The parameters $t_{i,j}$ ($t_{i,j}^{\text{SO}}$) set the amplitude of spin-conserving (spin-non-conserving) hopping between dots $i$ and $j$, where $\beta_{\uparrow,\downarrow}=\pm 1$.

The second term in \cref{eq:hamfull} describes the sensor dots and hopping between the sensors and the Kitaev chain,
\begin{align}
    H_\text{S}^{(n,s)} &= \sum_\sigma \varepsilon_{s \sigma} d_{s\sigma}^\dagger d_{s \sigma} +  \frac{U}{2} n_s (n_s-1) \\
    & + \delta_{s,L} \sum_\sigma \Big[ t_{L , 1} \, d_{L \sigma}^\dagger d_{1\sigma} + \beta_\sigma t_{L,1}^{\text{SO}} \, d_{L \sigma}^\dagger d_{1 \bar{\sigma}}  + \text{H.c.} \Big] \nonumber\\
    & + \delta_{s,R} \sum_\sigma \Big[ t_{n , R} \, d_{n \sigma}^\dagger d_{R \sigma} + \beta_\sigma t_{n , R}^{\text{SO}} \, d_{n \sigma}^\dagger d_{R \bar{\sigma}} + \text{H.c.} \Big],\nonumber
\end{align}
using similar notation as above.
In this case, $\varepsilon_{s\sigma} = \varepsilon_s + E_Z\,\delta_{\sigma,\downarrow}$ while the Kronecker delta functions $\delta_{s,L}$ and $\delta_{s,R}$ choose which sensor dot is included.

In the following, we will limit ourselves to the cases where the number of sites is either $n=2$ or $n=3$.
For the case with $n=2$, locations of the sweet spots are known for the model we use~\cite{tsintzis_creating_2022} and we can thus use this case as a test bed to assess the basic working of the tuning algorithm.
For the case with $n=3$, we expect a larger number of sweet spots, the locations of which are not known a priori.
This case will be used to investigate the robustness of the algorithm when operating in a higher-dimensional parameter space.

In an experiment, one would connect the chain to a source and drain lead (also indicated in Fig.~\ref{fig:setup}) and measure the differential conductance across the array of dots as a function of $\varepsilon_L$ or $\varepsilon_R$ and the bias voltage applied to the source and drain.
The location of the lowest-bias conductance peak then reveals the energy splitting between the lowest even and odd parity states of the array, and its dependence on the sensor dot potential contains information about the Majorana quality of the states on the outer sites of the Kitaev chain~\cite{souto_probing_2023,prada_measuring_2017,clarke_experimentally_2017}.
The deviation from the ideal case (zero splitting independent of $\varepsilon_{L}$ or $\varepsilon_R$) can then be quantified into a loss function used by the tuning algorithm to navigate parameter space, as we will detail in Sec.~\ref{sec:results}.
In our numerical simulations, it suffices to diagonalize $H^{(n,s)}$ and extract the even--odd ground state splitting directly from its eigenenergies as this corresponds to the location of the lowest-bias peak~\cite{souto_probing_2023}, which saves considerable computational overhead as compared to simulating a full differential-conductance measurement.

% END MODEL %

%%%%%%%%%%%%%%
% AUTOTUNING %
%%%%%%%%%%%%%%

To find the optimal set of tuning parameters $\varepsilon_{j,Sj}$ we use the CMA-ES algorithm \footnote{We use the \texttt{cma} Python library from \href{https://pypi.org/project/cma/}{pypi.org/project/cma}}, which is a simple evolution strategy optimization technique, relying on stochastic sampling rather than derivatives, making it suitable for problems where the objective function to be minimized for instance has several local minima or is non-smooth~\cite{hansen_completely_2001, hansen_reducing_2003, hansen_cma_2023}.
In our case, the limitation in sharpness of the conductance curves in real tunneling-spectroscopy measurements due to broadening from temperature and tunnel coupling to the leads \cite{kouwenhoven_electron_1997} was what motivated us to use a stochastic and derivative-free method: we can expect the limited resolution of an actual experiment to make it difficult to accurately evaluate derivatives of the loss function, likely rendering derivative-based unsuitable for optimization of the loss. Additionally, we are dealing with metrics containing several minima, where a purely derivative-based method may get lost in a non-optimal local minimum. 

\begin{figure}[t!]
    \centering
    \includegraphics[width=0.8\linewidth]{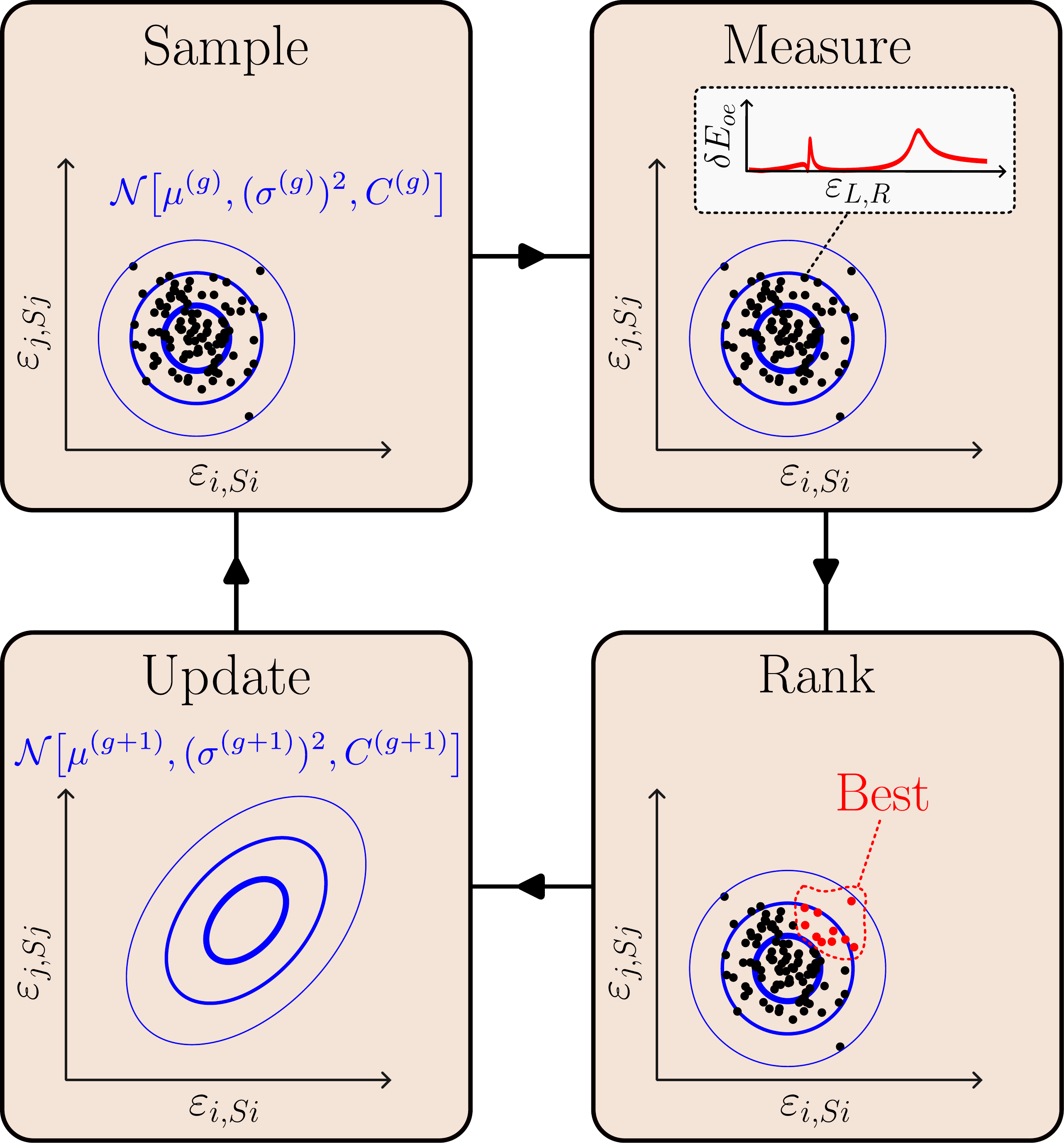}
    \caption{Diagram of the CMA-ES algorithm interfaced to either a simulation or an actual experiment where local differential conductances are measured. Each generation can be divided into four stages: The first stage is to sample $\lambda$ realizations from the MvN distribution to generate parameter configurations. In the second stage, each of these configurations is used in an experiment, where the dependence of the location of the low-bias conductance peaks (or the numerically determined lowest even--odd splittings) on the sensor dot voltage provides a measure for the population fitness. The third stage is to rank all the parameter configurations according to this fitness. In the fourth stage, the MvN mean value and covariance matrix are updated based on a subset of the $\eta$ best-ranked candidates.}
    \label{fig:cmaes}
\end{figure}

The CMA-ES algorithm performs the optimization by iterative sampling from and updating of a multivariate normal (MvN) distribution. For each generation $g$, a number of $\lambda \geq 2$ offspring $\boldsymbol\varepsilon^{(g)}$, which in this case are configurations of the gate voltages to tune, are randomly drawn from a MvN distribution with mean $\boldsymbol\mu^{(g)}$ and covariance matrix $\Sigma^{(g)} = (\sigma^{(g)})^2 C^{(g)}$. 
The configurations $\boldsymbol\varepsilon^{(g)}$ and means $\boldsymbol\mu^{(g)}$ are thus ($2n-1$)-dimensional vectors, e.g., for the two-site Kitaev chain $\boldsymbol\varepsilon^{(g)} = \{ \varepsilon_1^{(g)}, \varepsilon_{S1}^{(g)}, \varepsilon_2^{(g)} \}$.
The offspring are then ranked according to their ``fitness'' using an objective (loss) function ${\cal L}(\boldsymbol\varepsilon^{(g)})$, and the MvN distribution is updated by calculating a new position $\mu^{(g+1)}$, ``stepsize'' $(\sigma^{(g+1)})^2$ and ``search direction'' $C^{(g+1)}$, based on the $\eta$ best-ranked offspring and the previous history of changes to the population. 
A diagram of the CMA-ES optimization loop is shown in \cref{fig:cmaes}.
We will consistently use $\lambda = 56$ and $\eta = 28$, based mainly on the computational resources available, and set all other fine-tuning parameters of the algorithm to the standard recommended values~\cite{hansen_cma_2023}.
Specifically, the algorithm achieves convergence if (i) the range of best loss values across the last $10 + \lceil{30 (2n-1) /\lambda}\rceil$ generations or the range of all loss values in the last generation is smaller than $10^{-12}$ or (ii) if the standard deviation of all tuning parameters $\varepsilon_{j,Sj}$ becomes smaller than $10^{-12} \sigma^{(0)}$~\cite{hansen_cma_2023}. We also terminate the algorithm if it exceeds $200$ generations.

% END AUTOTUNING %

%%%%%%%%%%%
% RESULTS %
%%%%%%%%%%%

%\FloatBarrier
\section{Results}
\label{sec:results}

For the tuning of an artificial Kitaev chain, it has been suggested that tracking the splitting between the lowest conductance peaks while sweeping the sensor dot level $\varepsilon_s$ can provide a metric for the Majorana quality~\cite{souto_probing_2023}.
Following this idea, we will investigate the performance of the CMA-ES algorithm using the loss function
\begin{equation}
\label{max_esplit}
    f(\boldsymbol\varepsilon^{(g)}) = \frac{1}{2} \sum_{s=L,R} \max_{\varepsilon_{s}} \left|\frac{E_{0,n,s}^{\text{odd}} - E_{0,n,s}^{\text{even}}}{\Delta} \right|,
\end{equation}
where $E_{0,n,s}^{\text{odd}}$ and $E_{0,n,s}^{\text{even}}$ are the lowest eigenenergies of $H^{(n,s)}$ with total odd and even number of electrons, respectively.
The loss function $f(\boldsymbol\varepsilon^{(g)})$ thus returns the maximum of the absolute energy splitting between the even and odd ground states over a sweep of the sensor dot potential, averaged over the cases where the sensor dot is attached to the left or right side of the chain. The sweep range used for $\varepsilon_s$ should cover both spin-species resonances, the locations of which could, e.g., be found roughly from estimates of $E_Z$, $U$ and $\Delta$ \cite{prada_measuring_2017}.

In addition, we make use of an $L_2$ regularization to penalize on-site potentials that grow large, as the amplitudes of CAR and ECT are inversely proportional to the energy difference between the bound states on neighboring dots~\cite{liu_tunable_2022} and the magnitude of these amplitudes ultimately sets the scale of the topological gap in longer chains.
Combining $L_2$ regularization with the objective function \cref{max_esplit}, the final loss function we use to rank candidate parameter configurations is
\begin{equation}
\label{loss}
    \mathcal{L}(\boldsymbol\varepsilon^{(g)}) = f(\boldsymbol\varepsilon^{(g)}) + \alpha \bigg\lVert \frac{\boldsymbol\varepsilon^{(g)} }{\Delta} \bigg\rVert_2,
\end{equation}
where $\alpha$ parameterizes the $L_2$ regularization. 
For such regularization, one needs to establish some ``origin'' in parameter space, to determine $\lVert \boldsymbol\varepsilon^{(g)} \rVert_2$.
Identifying in an experiment the true origin where $\boldsymbol\varepsilon^{(g)}=0$ can be challenging, but small deviations from this point are not expected to pose an immediate problem as long as the origin used is not too far away from high-quality sweet spots.

Throughout, we will set all tunneling amplitudes to $t_{i,j} = 0.5\, \Delta$ and $t^{\text{SO}}_{i,j} = 0.1\, \Delta$, except for the ones describing the coupling to the sensor dot, where we use $t_{L,1} = t_{n,R} = 0.25\,\Delta$ and $t^{\text{SO}}_{L,1} = t^\text{SO}_{n,R} = 0.05\,\Delta$.
The Zeeman energy is set to $E_{Z}=1.5\,\Delta$ and the onsite Coulomb interaction strength for the normal dots to $U=5\,\Delta$.
These values are on purpose chosen the same as in Refs.~\cite{tsintzis_creating_2022, souto_probing_2023}, allowing for a straightforward comparison of the results.
The mean of the initial distribution for the CMA-ES algorithm is always set to the origin, with an initial covariance matrix of $C^{(0)} = \mathbb{I}$, and $\sigma^{(0)} = 0.5\, \Delta$.

In all instances, we let the algorithm search for the optimal set of tuning parameters until it converges, using the loss function given in Eq.~(\ref{loss}).
We then assess the resulting set of tuning parameters $\boldsymbol\varepsilon^\text{fin}$ by evaluating (i) the resulting even--odd ground state splitting in the Kitaev chain (without any sensor dot attached) $\delta E_{eo} = | E^\text{odd}_{0,n} - E^\text{even}_{0,n}|$, as well as (ii) the Majorana polarization (MP), given by~\cite{sedlmayr_visualizing_2015, sticlet_spin_2012, aksenov_strong_2020,tsintzis_roadmap_2023}
\begin{equation}
    M_j = \frac{\sum_\sigma (w_{j\sigma}^2-z_{j\sigma}^2)}{\sum_\sigma (w_{j\sigma}^2+z_{j\sigma}^2)},
\end{equation}
where
\begin{align}
    w_{j\sigma} = {} & {} \bra{O} d_{j\sigma} + d_{j\sigma}^\dagger \ket{E}, \\
    z_{j\sigma} = {} & {} \bra{O} d_{j\sigma} - d_{j\sigma}^\dagger \ket{E},
\end{align}
where $\ket{E}$ and $\ket{O}$ are the even and odd ground states of $H_{\text{Kit}}^{(n)}$, respectively, and $E^\text{even}_{0,n}$ and $E^\text{odd}_{0,n}$ the corresponding ground-state energies.
With this definition $-1 \leq M_j \leq 1$, and $|M_j|=1$ corresponds to having a well-behaved single Majorana mode localized at site $j$.
In Kitaev chains with finite Zeeman energy and Coulomb interactions, there exist in general no points in parameter space that have $\delta E_{eo} = 0$ and $|M_1| = |M_n| = 1$, which would indicate the existence of two ``perfect'' PMMs occupying the outer dots of the chain.
Instead, it is common to look for sweet spots that maximize $|M_{1,n}|$ on the manifold in parameter space where $\delta E_{eo}=0$~\cite{tsintzis_creating_2022,tsintzis_roadmap_2023}.
Below we will thus use $\delta E_{eo}$ and the average MP $M \equiv (|M_1|+|M_n|)/2$ as measures for the success of the tuning procedure.

\subsection{Two-site Kitaev chain}
\label{sec:twosite}

\begin{figure}[t]
    \centering
    \includegraphics[width=\linewidth]{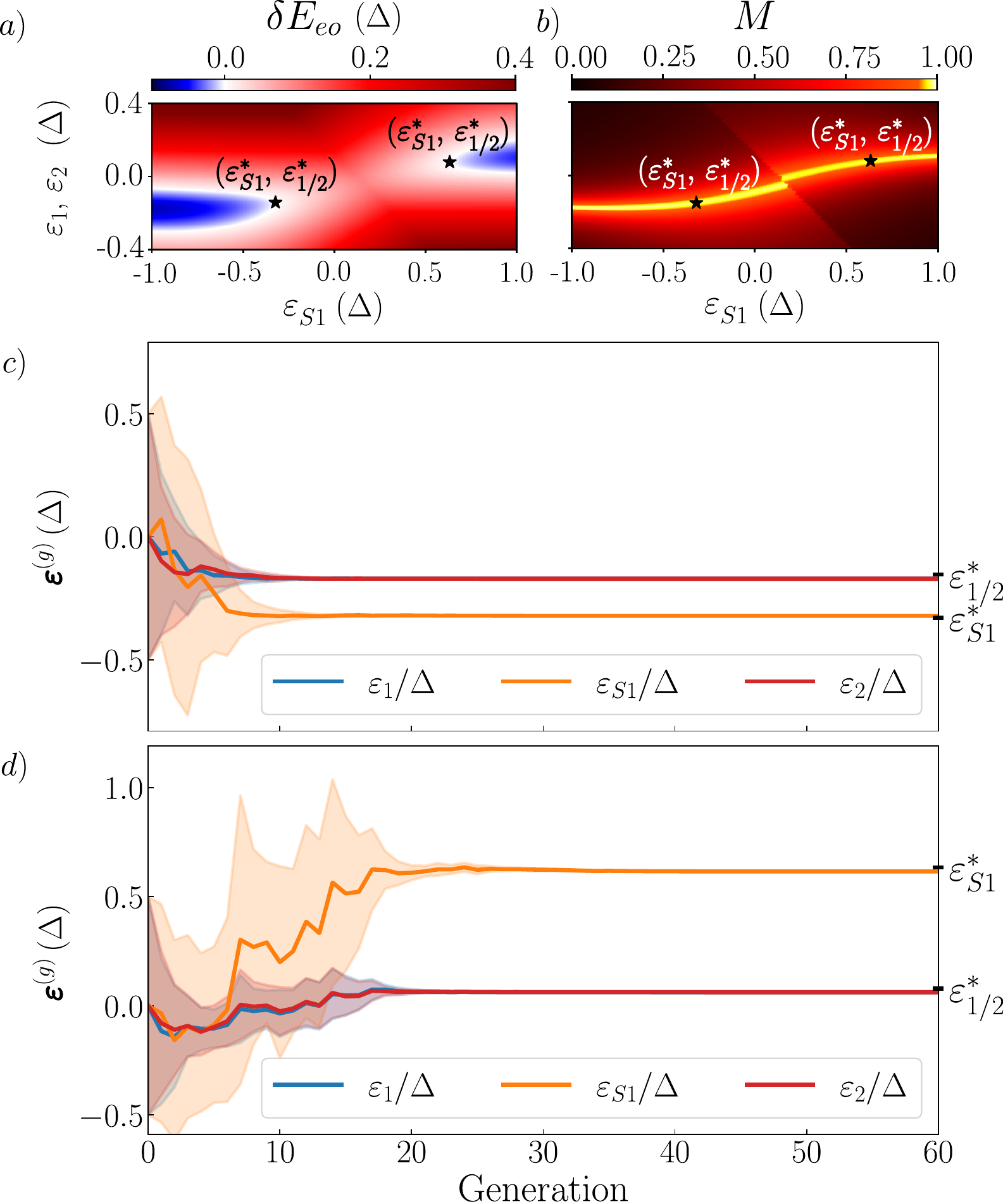}
    \caption{(a) The difference between even and odd ground state energy of the two-site Kitaev chain without a sensor dot attached and (b) the average Majorana polarization at its outer sites as a function of $\varepsilon_{1,2}$ and $\varepsilon_{S1}$, cf.~Ref.~\cite{tsintzis_creating_2022}.
    There are two sweet spots within the given parameter range, indicated by the black stars.
    (c,d) The $60$ first generations for two example runs of the CMA-ES algorithm for (c) a case that converged to the left sweet spot shown in (a,b) and (d) a case that found the right sweet spot.
    Solid lines show the population mean of the parameters and the shaded regions give their marginal standard deviation.}
    \label{fig:two_site_max_esplit}
\end{figure}

As a first benchmark, we test the optimization routine on a simulated minimal (two-site) Kitaev chain with $n=2$, where the sweet spots for our set of system parameters have been numerically calculated before~\cite{tsintzis_creating_2022}, to see if it can reproduce the same results.

In \cref{fig:two_site_max_esplit}(a,b) we show the even--odd ground state splitting $\delta E_{eo}$ and average polarization $M$ as a function of $\varepsilon_{1,2}$ and $\varepsilon_{S1}$, calculated numerically from $H_{\text{Kit}}^{(2)}$.
As expected, these plots are identical to the ones presented in Ref.~\cite{tsintzis_creating_2022}, since we use the same model and same parameters.
The two sweet spots identified in \cite{tsintzis_creating_2022} are indicated with the two stars, and have $\delta E_{eo} = 0$ and $M = 0.986$.

\cref{fig:two_site_max_esplit}(c,d) display the first $60$ generations of two example runs out of in total $50$ simulations for the CMA-ES optimization algorithm, using the loss function given in \cref{loss} while setting $\alpha = 0$. In the case of the minimal Kitaev chain, the objective function landscape is simple enough that $L_2$ regularization is unnecessary; including a finite $\alpha$ would increase the probability for the algorithm to discover the left sweet spot, as the right sweet spot is further away from the origin $\bm{\varepsilon}=0$ and therefore receives a larger punishment from the regularization.
The solid lines in \cref{fig:two_site_max_esplit}(c,d) show the evolution of the population mean of the three parameters and the shaded regions represent their marginal standard deviation.
We see that from generation to generation the standard deviations of the population decrease, eventually vanishing as the mean value approaches a minimum in the loss.

\cref{fig:two_site_max_esplit}(c) shows a case where the algorithm converged towards the left sweet spot shown in \cref{fig:two_site_max_esplit}(a,b), settling at the parameter configuration $\varepsilon_1 = \varepsilon_2 = -0.171\,\Delta$ and $\varepsilon_{S1} = -0.321\,\Delta$.
The ``target'' values, corresponding to the corresponding sweet spot identified in Ref.~\cite{tsintzis_creating_2022}, are indicated at the right vertical axis, where $\varepsilon_1^\ast = \varepsilon_2^\ast = -0.151\,\Delta$ and $\varepsilon_{S1}^\ast = -0.319\,\Delta$.
In $41$ cases out of $50$, the algorithm converged towards this sweet spot.
In the remaining $9$ cases, the algorithm converged towards the other sweet spot, an example of which is shown in \cref{fig:two_site_max_esplit}(d).
Here the final tuning parameters are $\varepsilon_1 = \varepsilon_2 = 0.0622\,\Delta$ and $\varepsilon_{S1} = 0.615\,\Delta$, which is again close the true location of the right sweet spot at $\varepsilon_1^\ast = \varepsilon_2^\ast = 0.0785\,\Delta$ and $\varepsilon_{S1}^\ast = 0.634\,\Delta$. The bias towards finding the left sweet spot is due to the initial MvN being centered around $\bm{\varepsilon}=0$, which is closer to the left sweet spot than the right one.
Convergence in these simulations was typically achieved after $80$--$100$ generations.

Thus we see that the algorithm always converges to tuning parameters that indeed lie very close to the sweet spots found in Ref.~\cite{tsintzis_creating_2022}, deviating from the sweet spots by Euclidian distances $\lVert \bm{\varepsilon} - \bm{\varepsilon^\ast} \rVert_2$ of $0.028 \,\Delta$ and $0.030 \,\Delta$ for the left and right sweet spots respectively. As a result, the left sweet spot estimate found by the algorithm has $\delta E_{eo} = 2.7 \times 10^{-3} \,\Delta$ and $M = 0.967$ while the right sweet spot estimate has $\delta E_{eo} = 5.7\times 10^{-3} \,\Delta$ and $M = 0.977$. 
In both cases we thus find high values for the MP and low splittings $\delta E_{eo}$~\footnote{Although such residual $\delta E_{eo}$ would still translate into a relatively short upper-bound time scale for ``Majorana manipulation,'' it lies well within the resolution of an actual tunnelling-spectroscopy experiment, likely making it impossible to resolve such small energy splitting anyway.
Furthermore, the automated tuning found here could serve as the starting point for a finer search using other methods.}, and
the two-site Kitaev chain results indicate that the very simple tunneling spectroscopy metric together with the CMA-ES algorithm can indeed tune on-site potentials $\bm{\varepsilon}$ to their sweet spots.

\subsection{Three-site Kitaev chain}
\label{sec:threesite}

To get an idea of how the automatic tuning will fare for longer chains, where the existence and locations of sweet spots are not known, we also tested it for the case of $n=3$, i.e., a five-dot array corresponding to a three-site Kitaev chain. We again run the optimization $50$ times, to obtain statistics on what solutions the algorithm tends to find. In this case we add an $L_2$ term with $\alpha=0.02$ as described by \cref{loss}.
The larger $\alpha$ is set, the more biased the algorithm is towards finding the sweet spot that is closest to the origin $\boldsymbol{\varepsilon} = 0$.
In an experiment, this parameter should be tuned empirically, where in order to determine a suitable $\alpha$ one could run the algorithm first without any regularization to find the typical order of magnitude for the objective function, and based on this, one can estimate how much regularization is needed to keep the gate voltages inside a given parameter range.

\begin{table*}
	\centering
	\begin{ruledtabular}
		\begin{tabular}{cccccc|ccc}
			&$\varepsilon_1/\Delta$& $\varepsilon_{S_1}/\Delta$ & $\varepsilon_2/\Delta$ & $\varepsilon_{S_2}/\Delta$ & $\varepsilon_3/\Delta$ & MP & $\delta E_{eo}/\Delta$ & Loss  \\ \hline
			$\bigtriangledown$ & $-0.10$ & $0.12$ & $0.14$ & $0.12$ & $-0.10$ & $0.998$ & $7.1\cdot 10^{-4}$ & $4.1\cdot 10^{-3}$  \\
			$\square$ & $0.19$ & $0.14$ & $-0.31$ & $-0.28$ & $-0.18$ & $0.979$ & $2.5\cdot 10^{-4}$ & $1.1\cdot 10^{-2}$  \\
			\protect\tikz{\draw[black] (0,0) circle (.8ex);} & $0.07$ & $0.20$ & $-0.40$ & $-0.32$ & $-0.18$ & $0.985$ & $8.7\cdot 10^{-4}$ & $1.2\cdot 10^{-2}$ 
		\end{tabular}
	\end{ruledtabular}
	\caption{(left) Overview of the three configurations of automatically tuned on-site potentials found in $50$ simulated experiments on a three-site Kitaev chain, using the loss function \cref{loss} with $\alpha=0.02$.
		(right) The corresponding Majorana polarization, even--odd ground state energy splitting, and final loss.
		The number of occurrences of each configuration is shown in the inset of \cref{fig:three_site_max_esplit}(a).}
	\label{tab:max_esplit}
\end{table*}

\begin{figure}[t]
    \centering
    \includegraphics[width=\linewidth]{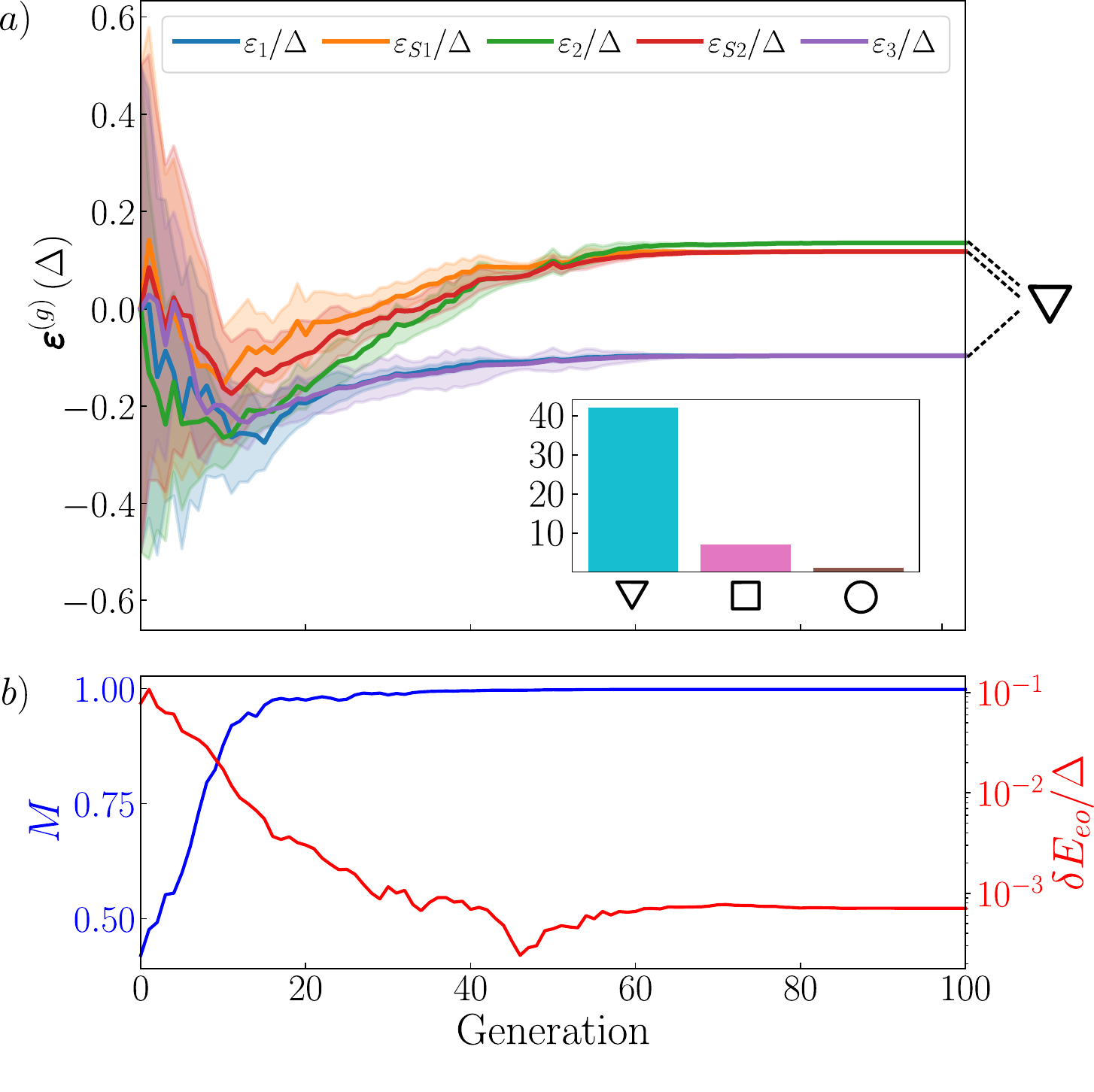}
    \caption{Development of the $100$ first generations for an example run of the CMA-ES algorithm for the three-site Kitaev chain, where it converges to the most commonly found parameter configuration. (a) Population mean of the parameters (solid lines) and their marginal standard deviation (shaded regions). (b) Majorana polarization $M$ (blue) and even--odd ground state splitting in the Kitaev chain $\delta E_{oe}$ (red) for each of the first $100$ generations.
    (Inset in a) Histogram of the occurrences of the three different sweet spots discovered in $50$ simulations.}
    \label{fig:three_site_max_esplit}
\end{figure}

The resulting automatically tuned potential configurations are summarized in \cref{tab:max_esplit} and a histogram of their occurrences is shown in the inset in \cref{fig:three_site_max_esplit}(a).
All of the simulations converge to configurations with MPs over $0.95$ and even--odd splittings below $10^{-3}\,\Delta$.
The exact sweet spot locations should have exactly zero $\delta E_{eo}$, and while we do see some finite value for all solutions, their small magnitude indicates that we get very close to the sweet spots.
We see that a large majority of the runs ($42$ out of $50$) ends up at a sweet spot with a MP of 0.998, which is significantly larger than at the two-site sweet spot.
Indeed, the addition of extra sites is expected to increase the maximal MPs that can be achieved, as compared to the minimal two-site chain.
%symmetrical parameter configuration, where 
To illustrate the working of the algorithm, we show the optimization process for one of these $42$ high-MP simulations in \cref{fig:three_site_max_esplit}.
In \cref{fig:three_site_max_esplit}(a) we show the CMA-ES population means and standard deviations for each parameter across generations and in \cref{fig:three_site_max_esplit}(b) we plot the corresponding values for the MP $M$ and energy splitting $\delta E_{eo}$.

We further note that for the 8 remaining cases the algorithm tuned the system to non-symmetric parameter configurations with high MP and low $\delta E_{eo}$, see \cref{tab:max_esplit}, which looks odd at first sight.
However, a feature of the three-site chain is that two pairs of ECT and CAR processes must be balanced rather than just one as in the minimal chain case.
The balanced pairs may differ from each other, allowing indeed for non-symmetrical configurations of the tuned on-site potentials along the chain.
In principle, these non-symmetric configurations could thus also correspond to valid sweet spots.
%In any case, these unsymmetrical outcomes seem to occur a lot less frequently than the symmetric ones as long as a sufficiently large $L_2$ regularization is applied. 

In the three-site chain simulations, convergence was achieved almost always within $140$--$200$ generations due to the change in the loss function becoming smaller than the set tolerance, two exceptions being experiments that terminated as a result of reaching the $200$-generation limit.
We finally note that, as one increases the system size to longer and longer chains (without drastically increasing $\lambda$), it can be expected that the algorithm will to a lesser extent find one sweet spot most of the time, and rather discover several different minima. This is a result of the sample statistics becoming more and more sparse as the dimensionality of the parameter space becomes large. In the opposite limit (small dimensionality and many samples $\lambda$), we expect the algorithm to become more and more deterministic, approaching a situation where one is essentially checking all solutions within the search area given by the MvN.

%%%%%%%%%%%%%%%
% END RESULTS %
%%%%%%%%%%%%%%%

%%%%%%%%%%%%%%
% CONCLUSION %
%%%%%%%%%%%%%%

%\FloatBarrier
\section{Conclusions}
\label{sec:conclusion}

We have simulated short artificial Kitaev chains based on arrays of quantum dots, through exact diagonalization of a simple model Hamiltonian, and we showed that tunneling spectroscopy in the presence of additional sensor dots at the ends of the chain could be used to automatically tune the system to a Majorana sweet spot, for which we specifically used the CMA-ES algorithm in this work.
We tested a simple metric based solely on the maximum difference between even and odd ground state energies as a function of the sensor dot voltages, suggested by Ref.~\cite{souto_probing_2023}, together with $L_2$ regularization on the gate voltages.
For a two-site chain, the algorithm was able to find the known sweet spots for this system, as identified in Ref.~\cite{tsintzis_creating_2022}.
In the case of a three-site chain, the algorithm was also capable of reliably tuning to apparent sweet spots, provided sufficient $L_2$ regularization was applied.
We conclude that a simultaneous automatic tuning of all on-site potentials using tunneling spectroscopy through sensor dots and the CMA-ES algorithm presents a promising way forward for automatizing the search for Majorana sweet spots in quantum-dot-based artificial Kitaev chains.

% END CONCLUSION %

\FloatBarrier

\section*{Acknowledgements}

We acknowledge funding via the QuantERA II Programme under Grant Agreement No.\ 101017733, the work being part of INTFELLES-Project No.~333990, which is funded by the Research Council of Norway (RCN).
This work was further funded by the Swedish Research Council under Grant Agreement No.~2020-03412, the Spanish CM Talento Program (Project No.~2022-T1/IND-24070), the Spanish Ministry of Science, Innovation, and Universities through Grant No.\ PID2022-140552NA-I00, the European Research Council (ERC) under the European Union’s Horizon 2020 research and innovation programme under Grant Agreement No.~856526, Nanolund, and the Horizon Europe Framework Program of the European Commission through the European Innovation Council Pathfinder Grant No.~101115315, QuKiT.
Simulations were performed on resources provided by the NTNU IDUN/EPIC computing cluster \cite{sjalander_epic_2022}, and we thank the NTNU HPC group for their technical support.

%%%%%%%%%%%%%%
% References %
%%%%%%%%%%%%%%



\begin{thebibliography}{60}%
\makeatletter
\providecommand \@ifxundefined [1]{%
 \@ifx{#1\undefined}
}%
\providecommand \@ifnum [1]{%
 \ifnum #1\expandafter \@firstoftwo
 \else \expandafter \@secondoftwo
 \fi
}%
\providecommand \@ifx [1]{%
 \ifx #1\expandafter \@firstoftwo
 \else \expandafter \@secondoftwo
 \fi
}%
\providecommand \natexlab [1]{#1}%
\providecommand \enquote  [1]{``#1''}%
\providecommand \bibnamefont  [1]{#1}%
\providecommand \bibfnamefont [1]{#1}%
\providecommand \citenamefont [1]{#1}%
\providecommand \href@noop [0]{\@secondoftwo}%
\providecommand \href [0]{\begingroup \@sanitize@url \@href}%
\providecommand \@href[1]{\@@startlink{#1}\@@href}%
\providecommand \@@href[1]{\endgroup#1\@@endlink}%
\providecommand \@sanitize@url [0]{\catcode `\\12\catcode `\$12\catcode
  `\&12\catcode `\#12\catcode `\^12\catcode `\_12\catcode `\%12\relax}%
\providecommand \@@startlink[1]{}%
\providecommand \@@endlink[0]{}%
\providecommand \url  [0]{\begingroup\@sanitize@url \@url }%
\providecommand \@url [1]{\endgroup\@href {#1}{\urlprefix }}%
\providecommand \urlprefix  [0]{URL }%
\providecommand \Eprint [0]{\href }%
\providecommand \doibase [0]{https://doi.org/}%
\providecommand \selectlanguage [0]{\@gobble}%
\providecommand \bibinfo  [0]{\@secondoftwo}%
\providecommand \bibfield  [0]{\@secondoftwo}%
\providecommand \translation [1]{[#1]}%
\providecommand \BibitemOpen [0]{}%
\providecommand \bibitemStop [0]{}%
\providecommand \bibitemNoStop [0]{.\EOS\space}%
\providecommand \EOS [0]{\spacefactor3000\relax}%
\providecommand \BibitemShut  [1]{\csname bibitem#1\endcsname}%
\let\auto@bib@innerbib\@empty
%</preamble>
\bibitem [{\citenamefont {Beenakker}(2013)}]{beenakker_search_2013}%
  \BibitemOpen
  \bibfield  {author} {\bibinfo {author} {\bibfnamefont {C.~W.~J.}\
  \bibnamefont {Beenakker}},\ }\bibfield  {title} {\bibinfo {title} {Search for
  {Majorana} {fermions} in {superconductors}},\ }\href
  {https://doi.org/10.1146/annurev-conmatphys-030212-184337} {\bibfield
  {journal} {\bibinfo  {journal} {Ann. Rev. Cond. Matt. Phys.}\ }\textbf
  {\bibinfo {volume} {4}},\ \bibinfo {pages} {113} (\bibinfo {year}
  {2013})}\BibitemShut {NoStop}%
\bibitem [{\citenamefont {Beenakker}(2020)}]{beenakker_search_2020}%
  \BibitemOpen
  \bibfield  {author} {\bibinfo {author} {\bibfnamefont {C.~W.~J.}\
  \bibnamefont {Beenakker}},\ }\bibfield  {title} {\bibinfo {title} {Search for
  non-{Abelian} {Majorana} braiding statistics in superconductors},\ }\href
  {https://doi.org/10.21468/SciPostPhysLectNotes.15} {\bibfield  {journal}
  {\bibinfo  {journal} {SciPost Phys. Lect. Notes},\ \bibinfo {pages} {15}}
  (\bibinfo {year} {2020})}\BibitemShut {NoStop}%
\bibitem [{\citenamefont {Aasen}\ \emph {et~al.}(2016)\citenamefont {Aasen},
  \citenamefont {Hell}, \citenamefont {Mishmash}, \citenamefont {Higginbotham},
  \citenamefont {Danon}, \citenamefont {Leijnse}, \citenamefont {Jespersen},
  \citenamefont {Folk}, \citenamefont {Marcus}, \citenamefont {Flensberg},\
  and\ \citenamefont {Alicea}}]{aasen_milestones_2016}%
  \BibitemOpen
  \bibfield  {author} {\bibinfo {author} {\bibfnamefont {D.}~\bibnamefont
  {Aasen}}, \bibinfo {author} {\bibfnamefont {M.}~\bibnamefont {Hell}},
  \bibinfo {author} {\bibfnamefont {R.~V.}\ \bibnamefont {Mishmash}}, \bibinfo
  {author} {\bibfnamefont {A.}~\bibnamefont {Higginbotham}}, \bibinfo {author}
  {\bibfnamefont {J.}~\bibnamefont {Danon}}, \bibinfo {author} {\bibfnamefont
  {M.}~\bibnamefont {Leijnse}}, \bibinfo {author} {\bibfnamefont {T.~S.}\
  \bibnamefont {Jespersen}}, \bibinfo {author} {\bibfnamefont {J.~A.}\
  \bibnamefont {Folk}}, \bibinfo {author} {\bibfnamefont {C.~M.}\ \bibnamefont
  {Marcus}}, \bibinfo {author} {\bibfnamefont {K.}~\bibnamefont {Flensberg}},\
  and\ \bibinfo {author} {\bibfnamefont {J.}~\bibnamefont {Alicea}},\
  }\bibfield  {title} {\bibinfo {title} {Milestones {toward} {Majorana}-{based}
  {quantum} {computing}},\ }\href {https://doi.org/10.1103/PhysRevX.6.031016}
  {\bibfield  {journal} {\bibinfo  {journal} {Phys. Rev. X}\ }\textbf {\bibinfo
  {volume} {6}},\ \bibinfo {pages} {031016} (\bibinfo {year}
  {2016})}\BibitemShut {NoStop}%
\bibitem [{\citenamefont {Alicea}(2012)}]{alicea_new_2012}%
  \BibitemOpen
  \bibfield  {author} {\bibinfo {author} {\bibfnamefont {J.}~\bibnamefont
  {Alicea}},\ }\bibfield  {title} {\bibinfo {title} {New directions in the
  pursuit of {Majorana} fermions in solid state systems},\ }\href
  {https://doi.org/10.1088/0034-4885/75/7/076501} {\bibfield  {journal}
  {\bibinfo  {journal} {Rep. Prog. Phys.}\ }\textbf {\bibinfo {volume} {75}},\
  \bibinfo {pages} {076501} (\bibinfo {year} {2012})}\BibitemShut {NoStop}%
\bibitem [{\citenamefont {Alicea}\ \emph {et~al.}(2011)\citenamefont {Alicea},
  \citenamefont {Oreg}, \citenamefont {Refael}, \citenamefont {von Oppen},\
  and\ \citenamefont {Fisher}}]{alicea_non-abelian_2011}%
  \BibitemOpen
  \bibfield  {author} {\bibinfo {author} {\bibfnamefont {J.}~\bibnamefont
  {Alicea}}, \bibinfo {author} {\bibfnamefont {Y.}~\bibnamefont {Oreg}},
  \bibinfo {author} {\bibfnamefont {G.}~\bibnamefont {Refael}}, \bibinfo
  {author} {\bibfnamefont {F.}~\bibnamefont {von Oppen}},\ and\ \bibinfo
  {author} {\bibfnamefont {M.~P.~A.}\ \bibnamefont {Fisher}},\ }\bibfield
  {title} {\bibinfo {title} {Non-{Abelian} statistics and topological quantum
  information processing in {1D} wire networks},\ }\href
  {https://doi.org/10.1038/nphys1915} {\bibfield  {journal} {\bibinfo
  {journal} {Nat. Phys.}\ }\textbf {\bibinfo {volume} {7}},\ \bibinfo {pages}
  {412} (\bibinfo {year} {2011})}\BibitemShut {NoStop}%
\bibitem [{\citenamefont {Nayak}\ \emph {et~al.}(2008)\citenamefont {Nayak},
  \citenamefont {Simon}, \citenamefont {Stern}, \citenamefont {Freedman},\ and\
  \citenamefont {Das~Sarma}}]{nayak_non-abelian_2008}%
  \BibitemOpen
  \bibfield  {author} {\bibinfo {author} {\bibfnamefont {C.}~\bibnamefont
  {Nayak}}, \bibinfo {author} {\bibfnamefont {S.~H.}\ \bibnamefont {Simon}},
  \bibinfo {author} {\bibfnamefont {A.}~\bibnamefont {Stern}}, \bibinfo
  {author} {\bibfnamefont {M.}~\bibnamefont {Freedman}},\ and\ \bibinfo
  {author} {\bibfnamefont {S.}~\bibnamefont {Das~Sarma}},\ }\bibfield  {title}
  {\bibinfo {title} {Non-{Abelian} anyons and topological quantum
  computation},\ }\href {https://doi.org/10.1103/RevModPhys.80.1083} {\bibfield
   {journal} {\bibinfo  {journal} {Rev. Mod. Phys.}\ }\textbf {\bibinfo
  {volume} {80}},\ \bibinfo {pages} {1083} (\bibinfo {year}
  {2008})}\BibitemShut {NoStop}%
\bibitem [{\citenamefont {Kitaev}(2001)}]{kitaev_unpaired_2001}%
  \BibitemOpen
  \bibfield  {author} {\bibinfo {author} {\bibfnamefont {A.~Y.}\ \bibnamefont
  {Kitaev}},\ }\bibfield  {title} {\bibinfo {title} {Unpaired {Majorana}
  fermions in quantum wires},\ }\href
  {https://doi.org/10.1070/1063-7869/44/10S/S29} {\bibfield  {journal}
  {\bibinfo  {journal} {Phys.-Usp.}\ }\textbf {\bibinfo {volume} {44}},\
  \bibinfo {pages} {131} (\bibinfo {year} {2001})}\BibitemShut {NoStop}%
\bibitem [{\citenamefont {Flensberg}\ \emph {et~al.}(2021)\citenamefont
  {Flensberg}, \citenamefont {von Oppen},\ and\ \citenamefont
  {Stern}}]{flensberg_engineered_2021}%
  \BibitemOpen
  \bibfield  {author} {\bibinfo {author} {\bibfnamefont {K.}~\bibnamefont
  {Flensberg}}, \bibinfo {author} {\bibfnamefont {F.}~\bibnamefont {von
  Oppen}},\ and\ \bibinfo {author} {\bibfnamefont {A.}~\bibnamefont {Stern}},\
  }\bibfield  {title} {\bibinfo {title} {Engineered platforms for topological
  superconductivity and {Majorana} zero modes},\ }\href
  {https://doi.org/10.1038/s41578-021-00336-6} {\bibfield  {journal} {\bibinfo
  {journal} {Nat. Rev. Mater.}\ }\textbf {\bibinfo {volume} {6}},\ \bibinfo
  {pages} {944} (\bibinfo {year} {2021})}\BibitemShut {NoStop}%
\bibitem [{\citenamefont {Deng}\ \emph {et~al.}(2016)\citenamefont {Deng},
  \citenamefont {Vaitiekėnas}, \citenamefont {Hansen}, \citenamefont {Danon},
  \citenamefont {Leijnse}, \citenamefont {Flensberg}, \citenamefont {Nygård},
  \citenamefont {Krogstrup},\ and\ \citenamefont
  {Marcus}}]{deng_majorana_2016}%
  \BibitemOpen
  \bibfield  {author} {\bibinfo {author} {\bibfnamefont {M.~T.}\ \bibnamefont
  {Deng}}, \bibinfo {author} {\bibfnamefont {S.}~\bibnamefont {Vaitiekėnas}},
  \bibinfo {author} {\bibfnamefont {E.~B.}\ \bibnamefont {Hansen}}, \bibinfo
  {author} {\bibfnamefont {J.}~\bibnamefont {Danon}}, \bibinfo {author}
  {\bibfnamefont {M.}~\bibnamefont {Leijnse}}, \bibinfo {author} {\bibfnamefont
  {K.}~\bibnamefont {Flensberg}}, \bibinfo {author} {\bibfnamefont
  {J.}~\bibnamefont {Nygård}}, \bibinfo {author} {\bibfnamefont
  {P.}~\bibnamefont {Krogstrup}},\ and\ \bibinfo {author} {\bibfnamefont
  {C.~M.}\ \bibnamefont {Marcus}},\ }\bibfield  {title} {\bibinfo {title}
  {Majorana bound state in a coupled quantum-dot hybrid-nanowire system},\
  }\href {https://doi.org/10.1126/science.aaf3961} {\bibfield  {journal}
  {\bibinfo  {journal} {Science}\ }\textbf {\bibinfo {volume} {354}},\ \bibinfo
  {pages} {1557} (\bibinfo {year} {2016})}\BibitemShut {NoStop}%
\bibitem [{\citenamefont {Deng}\ \emph {et~al.}(2012)\citenamefont {Deng},
  \citenamefont {Yu}, \citenamefont {Huang}, \citenamefont {Larsson},
  \citenamefont {Caroff},\ and\ \citenamefont {Xu}}]{deng_anomalous_2012}%
  \BibitemOpen
  \bibfield  {author} {\bibinfo {author} {\bibfnamefont {M.~T.}\ \bibnamefont
  {Deng}}, \bibinfo {author} {\bibfnamefont {C.~L.}\ \bibnamefont {Yu}},
  \bibinfo {author} {\bibfnamefont {G.~Y.}\ \bibnamefont {Huang}}, \bibinfo
  {author} {\bibfnamefont {M.}~\bibnamefont {Larsson}}, \bibinfo {author}
  {\bibfnamefont {P.}~\bibnamefont {Caroff}},\ and\ \bibinfo {author}
  {\bibfnamefont {H.~Q.}\ \bibnamefont {Xu}},\ }\bibfield  {title} {\bibinfo
  {title} {Anomalous zero-bias conductance peak in a {Nb}–{InSb}
  nanowire–{Nb} hybrid device},\ }\href {https://doi.org/10.1021/nl303758w}
  {\bibfield  {journal} {\bibinfo  {journal} {Nano Lett.}\ }\textbf {\bibinfo
  {volume} {12}},\ \bibinfo {pages} {6414} (\bibinfo {year}
  {2012})}\BibitemShut {NoStop}%
\bibitem [{\citenamefont {Finck}\ \emph {et~al.}(2013)\citenamefont {Finck},
  \citenamefont {Van~Harlingen}, \citenamefont {Mohseni}, \citenamefont
  {Jung},\ and\ \citenamefont {Li}}]{finck_anomalous_2013}%
  \BibitemOpen
  \bibfield  {author} {\bibinfo {author} {\bibfnamefont {A.~D.~K.}\
  \bibnamefont {Finck}}, \bibinfo {author} {\bibfnamefont {D.~J.}\ \bibnamefont
  {Van~Harlingen}}, \bibinfo {author} {\bibfnamefont {P.~K.}\ \bibnamefont
  {Mohseni}}, \bibinfo {author} {\bibfnamefont {K.}~\bibnamefont {Jung}},\ and\
  \bibinfo {author} {\bibfnamefont {X.}~\bibnamefont {Li}},\ }\bibfield
  {title} {\bibinfo {title} {Anomalous modulation of a zero-bias peak in a
  hybrid nanowire-superconductor device},\ }\href
  {https://doi.org/10.1103/PhysRevLett.110.126406} {\bibfield  {journal}
  {\bibinfo  {journal} {Phys. Rev. Lett.}\ }\textbf {\bibinfo {volume} {110}},\
  \bibinfo {pages} {126406} (\bibinfo {year} {2013})}\BibitemShut {NoStop}%
\bibitem [{\citenamefont {Lutchyn}\ \emph {et~al.}(2010)\citenamefont
  {Lutchyn}, \citenamefont {Sau},\ and\ \citenamefont
  {Das~Sarma}}]{lutchyn_majorana_2010}%
  \BibitemOpen
  \bibfield  {author} {\bibinfo {author} {\bibfnamefont {R.~M.}\ \bibnamefont
  {Lutchyn}}, \bibinfo {author} {\bibfnamefont {J.~D.}\ \bibnamefont {Sau}},\
  and\ \bibinfo {author} {\bibfnamefont {S.}~\bibnamefont {Das~Sarma}},\
  }\bibfield  {title} {\bibinfo {title} {Majorana {fermions} and a
  {topological} {phase} {transition} in {semiconductor}-{superconductor}
  {heterostructures}},\ }\href {https://doi.org/10.1103/PhysRevLett.105.077001}
  {\bibfield  {journal} {\bibinfo  {journal} {Phys. Rev. Lett.}\ }\textbf
  {\bibinfo {volume} {105}},\ \bibinfo {pages} {077001} (\bibinfo {year}
  {2010})}\BibitemShut {NoStop}%
\bibitem [{\citenamefont {Mourik}\ \emph {et~al.}(2012)\citenamefont {Mourik},
  \citenamefont {Zuo}, \citenamefont {Frolov}, \citenamefont {Plissard},
  \citenamefont {Bakkers},\ and\ \citenamefont
  {Kouwenhoven}}]{mourik_signatures_2012}%
  \BibitemOpen
  \bibfield  {author} {\bibinfo {author} {\bibfnamefont {V.}~\bibnamefont
  {Mourik}}, \bibinfo {author} {\bibfnamefont {K.}~\bibnamefont {Zuo}},
  \bibinfo {author} {\bibfnamefont {S.~M.}\ \bibnamefont {Frolov}}, \bibinfo
  {author} {\bibfnamefont {S.~R.}\ \bibnamefont {Plissard}}, \bibinfo {author}
  {\bibfnamefont {E.~P. A.~M.}\ \bibnamefont {Bakkers}},\ and\ \bibinfo
  {author} {\bibfnamefont {L.~P.}\ \bibnamefont {Kouwenhoven}},\ }\bibfield
  {title} {\bibinfo {title} {Signatures of {Majorana} {fermions} in {hybrid}
  {superconductor}-{semiconductor} {nanowire} {devices}},\ }\href
  {https://doi.org/10.1126/science.1222360} {\bibfield  {journal} {\bibinfo
  {journal} {Science}\ }\textbf {\bibinfo {volume} {336}},\ \bibinfo {pages}
  {1003} (\bibinfo {year} {2012})}\BibitemShut {NoStop}%
\bibitem [{\citenamefont {Nichele}\ \emph {et~al.}(2017)\citenamefont
  {Nichele}, \citenamefont {Drachmann}, \citenamefont {Whiticar}, \citenamefont
  {O’Farrell}, \citenamefont {Suominen}, \citenamefont {Fornieri},
  \citenamefont {Wang}, \citenamefont {Gardner}, \citenamefont {Thomas},
  \citenamefont {Hatke}, \citenamefont {Krogstrup}, \citenamefont {Manfra},
  \citenamefont {Flensberg},\ and\ \citenamefont
  {Marcus}}]{nichele_scaling_2017}%
  \BibitemOpen
  \bibfield  {author} {\bibinfo {author} {\bibfnamefont {F.}~\bibnamefont
  {Nichele}}, \bibinfo {author} {\bibfnamefont {A.~C.}\ \bibnamefont
  {Drachmann}}, \bibinfo {author} {\bibfnamefont {A.~M.}\ \bibnamefont
  {Whiticar}}, \bibinfo {author} {\bibfnamefont {E.~C.}\ \bibnamefont
  {O’Farrell}}, \bibinfo {author} {\bibfnamefont {H.~J.}\ \bibnamefont
  {Suominen}}, \bibinfo {author} {\bibfnamefont {A.}~\bibnamefont {Fornieri}},
  \bibinfo {author} {\bibfnamefont {T.}~\bibnamefont {Wang}}, \bibinfo {author}
  {\bibfnamefont {G.~C.}\ \bibnamefont {Gardner}}, \bibinfo {author}
  {\bibfnamefont {C.}~\bibnamefont {Thomas}}, \bibinfo {author} {\bibfnamefont
  {A.~T.}\ \bibnamefont {Hatke}}, \bibinfo {author} {\bibfnamefont
  {P.}~\bibnamefont {Krogstrup}}, \bibinfo {author} {\bibfnamefont {M.~J.}\
  \bibnamefont {Manfra}}, \bibinfo {author} {\bibfnamefont {K.}~\bibnamefont
  {Flensberg}},\ and\ \bibinfo {author} {\bibfnamefont {C.~M.}\ \bibnamefont
  {Marcus}},\ }\bibfield  {title} {\bibinfo {title} {Scaling of {Majorana}
  {zero}-{bias} {conductance} {peaks}},\ }\href
  {https://doi.org/10.1103/PhysRevLett.119.136803} {\bibfield  {journal}
  {\bibinfo  {journal} {Phys. Rev. Lett.}\ }\textbf {\bibinfo {volume} {119}},\
  \bibinfo {pages} {136803} (\bibinfo {year} {2017})}\BibitemShut {NoStop}%
\bibitem [{\citenamefont {Oreg}\ \emph {et~al.}(2010)\citenamefont {Oreg},
  \citenamefont {Refael},\ and\ \citenamefont {von Oppen}}]{oreg_helical_2010}%
  \BibitemOpen
  \bibfield  {author} {\bibinfo {author} {\bibfnamefont {Y.}~\bibnamefont
  {Oreg}}, \bibinfo {author} {\bibfnamefont {G.}~\bibnamefont {Refael}},\ and\
  \bibinfo {author} {\bibfnamefont {F.}~\bibnamefont {von Oppen}},\ }\bibfield
  {title} {\bibinfo {title} {Helical {liquids} and {Majorana} {bound} {states}
  in {quantum} {wires}},\ }\href
  {https://doi.org/10.1103/PhysRevLett.105.177002} {\bibfield  {journal}
  {\bibinfo  {journal} {Phys. Rev. Lett.}\ }\textbf {\bibinfo {volume} {105}},\
  \bibinfo {pages} {177002} (\bibinfo {year} {2010})}\BibitemShut {NoStop}%
\bibitem [{\citenamefont {Prada}\ \emph {et~al.}(2020)\citenamefont {Prada},
  \citenamefont {San-Jose}, \citenamefont {de~Moor}, \citenamefont {Geresdi},
  \citenamefont {Lee}, \citenamefont {Klinovaja}, \citenamefont {Loss},
  \citenamefont {Nygård}, \citenamefont {Aguado},\ and\ \citenamefont
  {Kouwenhoven}}]{prada_andreev_2020}%
  \BibitemOpen
  \bibfield  {author} {\bibinfo {author} {\bibfnamefont {E.}~\bibnamefont
  {Prada}}, \bibinfo {author} {\bibfnamefont {P.}~\bibnamefont {San-Jose}},
  \bibinfo {author} {\bibfnamefont {M.~W.~A.}\ \bibnamefont {de~Moor}},
  \bibinfo {author} {\bibfnamefont {A.}~\bibnamefont {Geresdi}}, \bibinfo
  {author} {\bibfnamefont {E.~J.~H.}\ \bibnamefont {Lee}}, \bibinfo {author}
  {\bibfnamefont {J.}~\bibnamefont {Klinovaja}}, \bibinfo {author}
  {\bibfnamefont {D.}~\bibnamefont {Loss}}, \bibinfo {author} {\bibfnamefont
  {J.}~\bibnamefont {Nygård}}, \bibinfo {author} {\bibfnamefont
  {R.}~\bibnamefont {Aguado}},\ and\ \bibinfo {author} {\bibfnamefont {L.~P.}\
  \bibnamefont {Kouwenhoven}},\ }\bibfield  {title} {\bibinfo {title} {From
  {Andreev} to {Majorana} bound states in hybrid superconductor–semiconductor
  nanowires},\ }\href {https://doi.org/10.1038/s42254-020-0228-y} {\bibfield
  {journal} {\bibinfo  {journal} {Nat. Rev. Phys.}\ }\textbf {\bibinfo {volume}
  {2}},\ \bibinfo {pages} {575} (\bibinfo {year} {2020})}\BibitemShut {NoStop}%
\bibitem [{\citenamefont {Liu}\ \emph {et~al.}(2012)\citenamefont {Liu},
  \citenamefont {Potter}, \citenamefont {Law},\ and\ \citenamefont
  {Lee}}]{liu_zero-bias_2012}%
  \BibitemOpen
  \bibfield  {author} {\bibinfo {author} {\bibfnamefont {J.}~\bibnamefont
  {Liu}}, \bibinfo {author} {\bibfnamefont {A.~C.}\ \bibnamefont {Potter}},
  \bibinfo {author} {\bibfnamefont {K.~T.}\ \bibnamefont {Law}},\ and\ \bibinfo
  {author} {\bibfnamefont {P.~A.}\ \bibnamefont {Lee}},\ }\bibfield  {title}
  {\bibinfo {title} {Zero-{bias} {peaks} in the {tunneling} {conductance} of
  {spin}-{orbit}-{coupled} {superconducting} {wires} with and without
  {Majorana} {end}-{states}},\ }\href
  {https://doi.org/10.1103/PhysRevLett.109.267002} {\bibfield  {journal}
  {\bibinfo  {journal} {Phys. Rev. Lett.}\ }\textbf {\bibinfo {volume} {109}},\
  \bibinfo {pages} {267002} (\bibinfo {year} {2012})}\BibitemShut {NoStop}%
\bibitem [{\citenamefont {Pikulin}\ \emph {et~al.}(2012)\citenamefont
  {Pikulin}, \citenamefont {Dahlhaus}, \citenamefont {Wimmer}, \citenamefont
  {Schomerus},\ and\ \citenamefont
  {Beenakker}}]{pikulinZerovoltageConductancePeak2012}%
  \BibitemOpen
  \bibfield  {author} {\bibinfo {author} {\bibfnamefont {D.~I.}\ \bibnamefont
  {Pikulin}}, \bibinfo {author} {\bibfnamefont {J.~P.}\ \bibnamefont
  {Dahlhaus}}, \bibinfo {author} {\bibfnamefont {M.}~\bibnamefont {Wimmer}},
  \bibinfo {author} {\bibfnamefont {H.}~\bibnamefont {Schomerus}},\ and\
  \bibinfo {author} {\bibfnamefont {C.~W.~J.}\ \bibnamefont {Beenakker}},\
  }\bibfield  {title} {\bibinfo {title} {A zero-voltage conductance peak from
  weak antilocalization in a {Majorana} nanowire},\ }\href
  {https://doi.org/10.1088/1367-2630/14/12/125011} {\bibfield  {journal}
  {\bibinfo  {journal} {New J. Phys.}\ }\textbf {\bibinfo {volume} {14}},\
  \bibinfo {pages} {125011} (\bibinfo {year} {2012})},\ \bibinfo {note}
  {publisher: IOP Publishing}\BibitemShut {NoStop}%
\bibitem [{\citenamefont {Kells}\ \emph {et~al.}(2012)\citenamefont {Kells},
  \citenamefont {Meidan},\ and\ \citenamefont
  {Brouwer}}]{kellsNearzeroenergyEndStates2012}%
  \BibitemOpen
  \bibfield  {author} {\bibinfo {author} {\bibfnamefont {G.}~\bibnamefont
  {Kells}}, \bibinfo {author} {\bibfnamefont {D.}~\bibnamefont {Meidan}},\ and\
  \bibinfo {author} {\bibfnamefont {P.~W.}\ \bibnamefont {Brouwer}},\
  }\bibfield  {title} {\bibinfo {title} {Near-zero-energy end states in
  topologically trivial spin-orbit coupled superconducting nanowires with a
  smooth confinement},\ }\href {https://doi.org/10.1103/physrevb.86.100503}
  {\bibfield  {journal} {\bibinfo  {journal} {Phys. Rev. B}\ }\textbf {\bibinfo
  {volume} {86}},\ \bibinfo {pages} {100503} (\bibinfo {year}
  {2012})}\BibitemShut {NoStop}%
\bibitem [{\citenamefont {Zhang}\ \emph {et~al.}(2017)\citenamefont {Zhang},
  \citenamefont {G{ü}l}, \citenamefont {Conesa-Boj}, \citenamefont {Nowak},
  \citenamefont {Wimmer}, \citenamefont {Zuo}, \citenamefont {Mourik},
  \citenamefont {de~Vries}, \citenamefont {van Veen}, \citenamefont {de~Moor},
  \citenamefont {Bommer}, \citenamefont {van Woerkom}, \citenamefont {Car},
  \citenamefont {Plissard}, \citenamefont {Bakkers}, \citenamefont
  {Quintero-Pérez}, \citenamefont {Cassidy}, \citenamefont {Koelling},
  \citenamefont {Goswami}, \citenamefont {Watanabe}, \citenamefont
  {Taniguchi},\ and\ \citenamefont {Kouwenhoven}}]{zhang_ballistic_2017}%
  \BibitemOpen
  \bibfield  {author} {\bibinfo {author} {\bibfnamefont {H.}~\bibnamefont
  {Zhang}}, \bibinfo {author} {\bibfnamefont {{\"O}.}~\bibnamefont {G{ü}l}},
  \bibinfo {author} {\bibfnamefont {S.}~\bibnamefont {Conesa-Boj}}, \bibinfo
  {author} {\bibfnamefont {M.~P.}\ \bibnamefont {Nowak}}, \bibinfo {author}
  {\bibfnamefont {M.}~\bibnamefont {Wimmer}}, \bibinfo {author} {\bibfnamefont
  {K.}~\bibnamefont {Zuo}}, \bibinfo {author} {\bibfnamefont {V.}~\bibnamefont
  {Mourik}}, \bibinfo {author} {\bibfnamefont {F.~K.}\ \bibnamefont
  {de~Vries}}, \bibinfo {author} {\bibfnamefont {J.}~\bibnamefont {van Veen}},
  \bibinfo {author} {\bibfnamefont {M.~W.~A.}\ \bibnamefont {de~Moor}},
  \bibinfo {author} {\bibfnamefont {J.~D.~S.}\ \bibnamefont {Bommer}}, \bibinfo
  {author} {\bibfnamefont {D.~J.}\ \bibnamefont {van Woerkom}}, \bibinfo
  {author} {\bibfnamefont {D.}~\bibnamefont {Car}}, \bibinfo {author}
  {\bibfnamefont {S.~R.}\ \bibnamefont {Plissard}}, \bibinfo {author}
  {\bibfnamefont {E.~P. A.~M.}\ \bibnamefont {Bakkers}}, \bibinfo {author}
  {\bibfnamefont {M.}~\bibnamefont {Quintero-Pérez}}, \bibinfo {author}
  {\bibfnamefont {M.~C.}\ \bibnamefont {Cassidy}}, \bibinfo {author}
  {\bibfnamefont {S.}~\bibnamefont {Koelling}}, \bibinfo {author}
  {\bibfnamefont {S.}~\bibnamefont {Goswami}}, \bibinfo {author} {\bibfnamefont
  {K.}~\bibnamefont {Watanabe}}, \bibinfo {author} {\bibfnamefont
  {T.}~\bibnamefont {Taniguchi}},\ and\ \bibinfo {author} {\bibfnamefont
  {L.~P.}\ \bibnamefont {Kouwenhoven}},\ }\bibfield  {title} {\bibinfo {title}
  {Ballistic superconductivity in semiconductor nanowires},\ }\href
  {https://doi.org/10.1038/ncomms16025} {\bibfield  {journal} {\bibinfo
  {journal} {Nat. Commun.}\ }\textbf {\bibinfo {volume} {8}},\ \bibinfo {pages}
  {16025} (\bibinfo {year} {2017})}\BibitemShut {NoStop}%
\bibitem [{\citenamefont {Reeg}\ \emph {et~al.}(2018)\citenamefont {Reeg},
  \citenamefont {Dmytruk}, \citenamefont {Chevallier}, \citenamefont {Loss},\
  and\ \citenamefont {Klinovaja}}]{reegZeroenergyAndreevBound2018}%
  \BibitemOpen
  \bibfield  {author} {\bibinfo {author} {\bibfnamefont {C.}~\bibnamefont
  {Reeg}}, \bibinfo {author} {\bibfnamefont {O.}~\bibnamefont {Dmytruk}},
  \bibinfo {author} {\bibfnamefont {D.}~\bibnamefont {Chevallier}}, \bibinfo
  {author} {\bibfnamefont {D.}~\bibnamefont {Loss}},\ and\ \bibinfo {author}
  {\bibfnamefont {J.}~\bibnamefont {Klinovaja}},\ }\bibfield  {title} {\bibinfo
  {title} {Zero-energy {Andreev} bound states from quantum dots in proximitized
  {Rashba} nanowires},\ }\href {https://doi.org/10.1103/physrevb.98.245407}
  {\bibfield  {journal} {\bibinfo  {journal} {Phys. Rev. B}\ }\textbf {\bibinfo
  {volume} {98}},\ \bibinfo {pages} {245407} (\bibinfo {year}
  {2018})}\BibitemShut {NoStop}%
\bibitem [{\citenamefont {Woods}\ \emph {et~al.}(2019)\citenamefont {Woods},
  \citenamefont {Chen}, \citenamefont {Frolov},\ and\ \citenamefont
  {Stanescu}}]{woodsZeroenergypinning2019}%
  \BibitemOpen
  \bibfield  {author} {\bibinfo {author} {\bibfnamefont {B.~D.}\ \bibnamefont
  {Woods}}, \bibinfo {author} {\bibfnamefont {J.}~\bibnamefont {Chen}},
  \bibinfo {author} {\bibfnamefont {S.~M.}\ \bibnamefont {Frolov}},\ and\
  \bibinfo {author} {\bibfnamefont {T.~D.}\ \bibnamefont {Stanescu}},\
  }\bibfield  {title} {\bibinfo {title} {Zero-energy pinning of topologically
  trivial bound states in multiband semiconductor-superconductor nanowires},\
  }\href {https://doi.org/10.1103/PhysRevB.100.125407} {\bibfield  {journal}
  {\bibinfo  {journal} {Phys. Rev. B}\ }\textbf {\bibinfo {volume} {100}},\
  \bibinfo {pages} {125407} (\bibinfo {year} {2019})}\BibitemShut {NoStop}%
\bibitem [{\citenamefont {Pan}\ and\ \citenamefont
  {Das~Sarma}(2020)}]{pan_physical_2020}%
  \BibitemOpen
  \bibfield  {author} {\bibinfo {author} {\bibfnamefont {H.}~\bibnamefont
  {Pan}}\ and\ \bibinfo {author} {\bibfnamefont {S.}~\bibnamefont
  {Das~Sarma}},\ }\bibfield  {title} {\bibinfo {title} {Physical mechanisms for
  zero-bias conductance peaks in {Majorana} nanowires},\ }\href
  {https://doi.org/10.1103/PhysRevResearch.2.013377} {\bibfield  {journal}
  {\bibinfo  {journal} {Phys. Rev. Research}\ }\textbf {\bibinfo {volume}
  {2}},\ \bibinfo {pages} {013377} (\bibinfo {year} {2020})}\BibitemShut
  {NoStop}%
\bibitem [{\citenamefont {Ahn}\ \emph {et~al.}(2021)\citenamefont {Ahn},
  \citenamefont {Pan}, \citenamefont {Woods}, \citenamefont {Stanescu},\ and\
  \citenamefont {Das~Sarma}}]{ahn_estimating_2021}%
  \BibitemOpen
  \bibfield  {author} {\bibinfo {author} {\bibfnamefont {S.}~\bibnamefont
  {Ahn}}, \bibinfo {author} {\bibfnamefont {H.}~\bibnamefont {Pan}}, \bibinfo
  {author} {\bibfnamefont {B.}~\bibnamefont {Woods}}, \bibinfo {author}
  {\bibfnamefont {T.~D.}\ \bibnamefont {Stanescu}},\ and\ \bibinfo {author}
  {\bibfnamefont {S.}~\bibnamefont {Das~Sarma}},\ }\bibfield  {title} {\bibinfo
  {title} {Estimating disorder and its adverse effects in semiconductor
  {Majorana} nanowires},\ }\href
  {https://doi.org/10.1103/PhysRevMaterials.5.124602} {\bibfield  {journal}
  {\bibinfo  {journal} {Phys. Rev. Materials}\ }\textbf {\bibinfo {volume}
  {5}},\ \bibinfo {pages} {124602} (\bibinfo {year} {2021})}\BibitemShut
  {NoStop}%
\bibitem [{\citenamefont {Das~Sarma}\ and\ \citenamefont
  {Pan}(2021)}]{das_sarma_disorder-induced_2021}%
  \BibitemOpen
  \bibfield  {author} {\bibinfo {author} {\bibfnamefont {S.}~\bibnamefont
  {Das~Sarma}}\ and\ \bibinfo {author} {\bibfnamefont {H.}~\bibnamefont
  {Pan}},\ }\bibfield  {title} {\bibinfo {title} {Disorder-induced zero-bias
  peaks in {Majorana} nanowires},\ }\href
  {https://doi.org/10.1103/PhysRevB.103.195158} {\bibfield  {journal} {\bibinfo
   {journal} {Phys. Rev. B}\ }\textbf {\bibinfo {volume} {103}},\ \bibinfo
  {pages} {195158} (\bibinfo {year} {2021})}\BibitemShut {NoStop}%
\bibitem [{\citenamefont {Yu}\ \emph {et~al.}(2021)\citenamefont {Yu},
  \citenamefont {Chen}, \citenamefont {Gomanko}, \citenamefont {Badawy},
  \citenamefont {Bakkers}, \citenamefont {Zuo}, \citenamefont {Mourik},\ and\
  \citenamefont {Frolov}}]{yu_non-majorana_2021}%
  \BibitemOpen
  \bibfield  {author} {\bibinfo {author} {\bibfnamefont {P.}~\bibnamefont
  {Yu}}, \bibinfo {author} {\bibfnamefont {J.}~\bibnamefont {Chen}}, \bibinfo
  {author} {\bibfnamefont {M.}~\bibnamefont {Gomanko}}, \bibinfo {author}
  {\bibfnamefont {G.}~\bibnamefont {Badawy}}, \bibinfo {author} {\bibfnamefont
  {E.~P. A.~M.}\ \bibnamefont {Bakkers}}, \bibinfo {author} {\bibfnamefont
  {K.}~\bibnamefont {Zuo}}, \bibinfo {author} {\bibfnamefont {V.}~\bibnamefont
  {Mourik}},\ and\ \bibinfo {author} {\bibfnamefont {S.~M.}\ \bibnamefont
  {Frolov}},\ }\bibfield  {title} {\bibinfo {title} {Non-{Majorana} states
  yield nearly quantized conductance in proximatized nanowires},\ }\href
  {https://doi.org/10.1038/s41567-020-01107-w} {\bibfield  {journal} {\bibinfo
  {journal} {Nat. Phys.}\ }\textbf {\bibinfo {volume} {17}},\ \bibinfo {pages}
  {482} (\bibinfo {year} {2021})}\BibitemShut {NoStop}%
\bibitem [{\citenamefont {Hess}\ \emph {et~al.}(2021)\citenamefont {Hess},
  \citenamefont {Legg}, \citenamefont {Loss},\ and\ \citenamefont
  {Klinovaja}}]{hessLocalNonlocalQuantum2021a}%
  \BibitemOpen
  \bibfield  {author} {\bibinfo {author} {\bibfnamefont {R.}~\bibnamefont
  {Hess}}, \bibinfo {author} {\bibfnamefont {H.~F.}\ \bibnamefont {Legg}},
  \bibinfo {author} {\bibfnamefont {D.}~\bibnamefont {Loss}},\ and\ \bibinfo
  {author} {\bibfnamefont {J.}~\bibnamefont {Klinovaja}},\ }\bibfield  {title}
  {\bibinfo {title} {Local and nonlocal quantum transport due to {Andreev}
  bound states in finite {Rashba} nanowires with superconducting and normal
  sections},\ }\href {https://doi.org/10.1103/PhysRevB.104.075405} {\bibfield
  {journal} {\bibinfo  {journal} {Phys. Rev. B}\ }\textbf {\bibinfo {volume}
  {104}},\ \bibinfo {pages} {075405} (\bibinfo {year} {2021})}\BibitemShut
  {NoStop}%
\bibitem [{\citenamefont {Cayao}\ and\ \citenamefont
  {Burset}(2021)}]{cayaoConfinementinducedZerobiasPeaks2021}%
  \BibitemOpen
  \bibfield  {author} {\bibinfo {author} {\bibfnamefont {J.}~\bibnamefont
  {Cayao}}\ and\ \bibinfo {author} {\bibfnamefont {P.}~\bibnamefont {Burset}},\
  }\bibfield  {title} {\bibinfo {title} {Confinement-induced zero-bias peaks in
  conventional superconductor hybrids},\ }\href
  {https://doi.org/10.1103/PhysRevB.104.134507} {\bibfield  {journal} {\bibinfo
   {journal} {Phys. Rev. B}\ }\textbf {\bibinfo {volume} {104}},\ \bibinfo
  {pages} {134507} (\bibinfo {year} {2021})}\BibitemShut {NoStop}%
\bibitem [{\citenamefont {Hess}\ \emph {et~al.}(2023)\citenamefont {Hess},
  \citenamefont {Legg}, \citenamefont {Loss},\ and\ \citenamefont
  {Klinovaja}}]{hessTrivialAndreevband2023}%
  \BibitemOpen
  \bibfield  {author} {\bibinfo {author} {\bibfnamefont {R.}~\bibnamefont
  {Hess}}, \bibinfo {author} {\bibfnamefont {H.~F.}\ \bibnamefont {Legg}},
  \bibinfo {author} {\bibfnamefont {D.}~\bibnamefont {Loss}},\ and\ \bibinfo
  {author} {\bibfnamefont {J.}~\bibnamefont {Klinovaja}},\ }\bibfield  {title}
  {\bibinfo {title} {Trivial {Andreev} band mimicking topological bulk gap
  reopening in the nonlocal conductance of long {Rashba} nanowires},\ }\href
  {https://doi.org/10.1103/PhysRevLett.130.207001} {\bibfield  {journal}
  {\bibinfo  {journal} {Phys. Rev. Lett.}\ }\textbf {\bibinfo {volume} {130}},\
  \bibinfo {pages} {207001} (\bibinfo {year} {2023})}\BibitemShut {NoStop}%
\bibitem [{\citenamefont {Sau}\ and\ \citenamefont
  {Sarma}(2012)}]{sau_realizing_2012}%
  \BibitemOpen
  \bibfield  {author} {\bibinfo {author} {\bibfnamefont {J.~D.}\ \bibnamefont
  {Sau}}\ and\ \bibinfo {author} {\bibfnamefont {S.~D.}\ \bibnamefont
  {Sarma}},\ }\bibfield  {title} {\bibinfo {title} {Realizing a robust
  practical {Majorana} chain in a quantum-dot-superconductor linear array},\
  }\href {https://doi.org/10.1038/ncomms1966} {\bibfield  {journal} {\bibinfo
  {journal} {Nat. Commun.}\ }\textbf {\bibinfo {volume} {3}},\ \bibinfo {pages}
  {964} (\bibinfo {year} {2012})}\BibitemShut {NoStop}%
\bibitem [{\citenamefont {Leijnse}\ and\ \citenamefont
  {Flensberg}(2012)}]{leijnse_parity_2012}%
  \BibitemOpen
  \bibfield  {author} {\bibinfo {author} {\bibfnamefont {M.}~\bibnamefont
  {Leijnse}}\ and\ \bibinfo {author} {\bibfnamefont {K.}~\bibnamefont
  {Flensberg}},\ }\bibfield  {title} {\bibinfo {title} {Parity qubits and poor
  man's {Majorana} bound states in double quantum dots},\ }\href
  {https://doi.org/10.1103/PhysRevB.86.134528} {\bibfield  {journal} {\bibinfo
  {journal} {Phys. Rev. B}\ }\textbf {\bibinfo {volume} {86}},\ \bibinfo
  {pages} {134528} (\bibinfo {year} {2012})}\BibitemShut {NoStop}%
\bibitem [{\citenamefont {Miles}\ \emph {et~al.}(2023)\citenamefont {Miles},
  \citenamefont {van Driel}, \citenamefont {Wimmer},\ and\ \citenamefont
  {Liu}}]{miles_kitaev_2023}%
  \BibitemOpen
  \bibfield  {author} {\bibinfo {author} {\bibfnamefont {S.}~\bibnamefont
  {Miles}}, \bibinfo {author} {\bibfnamefont {D.}~\bibnamefont {van Driel}},
  \bibinfo {author} {\bibfnamefont {M.}~\bibnamefont {Wimmer}},\ and\ \bibinfo
  {author} {\bibfnamefont {C.-X.}\ \bibnamefont {Liu}},\ }\bibfield  {title}
  {\bibinfo {title} {Kitaev chain in an alternating quantum dot-{Andreev} bound
  state array},\ }\href {http://arxiv.org/abs/2309.15777} {\bibfield  {journal}
  {\bibinfo  {journal} {arXiv:2309.15777}} (\bibinfo {year}
  {2023})}\BibitemShut {NoStop}%
\bibitem [{\citenamefont {Tsintzis}\ \emph {et~al.}(2022)\citenamefont
  {Tsintzis}, \citenamefont {Souto},\ and\ \citenamefont
  {Leijnse}}]{tsintzis_creating_2022}%
  \BibitemOpen
  \bibfield  {author} {\bibinfo {author} {\bibfnamefont {A.}~\bibnamefont
  {Tsintzis}}, \bibinfo {author} {\bibfnamefont {R.~S.}\ \bibnamefont
  {Souto}},\ and\ \bibinfo {author} {\bibfnamefont {M.}~\bibnamefont
  {Leijnse}},\ }\bibfield  {title} {\bibinfo {title} {Creating and detecting
  poor man's {Majorana} bound states in interacting quantum dots},\ }\href
  {https://doi.org/10.1103/PhysRevB.106.L201404} {\bibfield  {journal}
  {\bibinfo  {journal} {Phys. Rev. B}\ }\textbf {\bibinfo {volume} {106}},\
  \bibinfo {pages} {L201404} (\bibinfo {year} {2022})}\BibitemShut {NoStop}%
\bibitem [{\citenamefont {Tsintzis}\ \emph {et~al.}(2024)\citenamefont
  {Tsintzis}, \citenamefont {Souto}, \citenamefont {Flensberg}, \citenamefont
  {Danon},\ and\ \citenamefont {Leijnse}}]{tsintzis_roadmap_2023}%
  \BibitemOpen
  \bibfield  {author} {\bibinfo {author} {\bibfnamefont {A.}~\bibnamefont
  {Tsintzis}}, \bibinfo {author} {\bibfnamefont {R.~S.}\ \bibnamefont {Souto}},
  \bibinfo {author} {\bibfnamefont {K.}~\bibnamefont {Flensberg}}, \bibinfo
  {author} {\bibfnamefont {J.}~\bibnamefont {Danon}},\ and\ \bibinfo {author}
  {\bibfnamefont {M.}~\bibnamefont {Leijnse}},\ }\bibfield  {title} {\bibinfo
  {title} {Majorana qubits and non-abelian physics in quantum dot--based
  minimal kitaev chains},\ }\href {https://doi.org/10.1103/PRXQuantum.5.010323}
  {\bibfield  {journal} {\bibinfo  {journal} {PRX Quantum}\ }\textbf {\bibinfo
  {volume} {5}},\ \bibinfo {pages} {010323} (\bibinfo {year}
  {2024})}\BibitemShut {NoStop}%
\bibitem [{\citenamefont {Luna}\ \emph {et~al.}(2024)\citenamefont {Luna},
  \citenamefont {Bozkurt}, \citenamefont {Wimmer},\ and\ \citenamefont
  {Liu}}]{luna_flux-tunable_2024}%
  \BibitemOpen
  \bibfield  {author} {\bibinfo {author} {\bibfnamefont {J.~D.~T.}\
  \bibnamefont {Luna}}, \bibinfo {author} {\bibfnamefont {A.~M.}\ \bibnamefont
  {Bozkurt}}, \bibinfo {author} {\bibfnamefont {M.}~\bibnamefont {Wimmer}},\
  and\ \bibinfo {author} {\bibfnamefont {C.-X.}\ \bibnamefont {Liu}},\
  }\bibfield  {title} {\bibinfo {title} {Flux-tunable {Kitaev} chain in a
  quantum dot array},\ }\href {http://arxiv.org/abs/2402.07575} {\bibfield
  {journal} {\bibinfo  {journal} {arXiv:2402.07575}} (\bibinfo {year}
  {2024})}\BibitemShut {NoStop}%
\bibitem [{\citenamefont {Liu}\ \emph {et~al.}(2023{\natexlab{a}})\citenamefont
  {Liu}, \citenamefont {Pan}, \citenamefont {Setiawan}, \citenamefont
  {Wimmer},\ and\ \citenamefont {Sau}}]{liu_fusion_2023}%
  \BibitemOpen
  \bibfield  {author} {\bibinfo {author} {\bibfnamefont {C.-X.}\ \bibnamefont
  {Liu}}, \bibinfo {author} {\bibfnamefont {H.}~\bibnamefont {Pan}}, \bibinfo
  {author} {\bibfnamefont {F.}~\bibnamefont {Setiawan}}, \bibinfo {author}
  {\bibfnamefont {M.}~\bibnamefont {Wimmer}},\ and\ \bibinfo {author}
  {\bibfnamefont {J.~D.}\ \bibnamefont {Sau}},\ }\bibfield  {title} {\bibinfo
  {title} {Fusion protocol for {Majorana} modes in coupled quantum dots},\
  }\href {https://doi.org/10.1103/PhysRevB.108.085437} {\bibfield  {journal}
  {\bibinfo  {journal} {Phys. Rev. B}\ }\textbf {\bibinfo {volume} {108}},\
  \bibinfo {pages} {085437} (\bibinfo {year} {2023}{\natexlab{a}})}\BibitemShut
  {NoStop}%
\bibitem [{\citenamefont {Souto}\ and\ \citenamefont
  {Aguado}(2024)}]{souto_subgap_2024}%
  \BibitemOpen
  \bibfield  {author} {\bibinfo {author} {\bibfnamefont {R.~S.}\ \bibnamefont
  {Souto}}\ and\ \bibinfo {author} {\bibfnamefont {R.}~\bibnamefont {Aguado}},\
  }\bibfield  {title} {\bibinfo {title} {Subgap states in
  semiconductor-superconductor devices for quantum technologies: {Andreev}
  qubits and minimal {Majorana} chains},\ }\href
  {http://arxiv.org/abs/2404.06592} {\bibfield  {journal} {\bibinfo  {journal}
  {arXiv:2404.06592}} (\bibinfo {year} {2024})}\BibitemShut {NoStop}%
\bibitem [{\citenamefont {Dvir}\ \emph {et~al.}(2023)\citenamefont {Dvir},
  \citenamefont {Wang}, \citenamefont {van Loo}, \citenamefont {Liu},
  \citenamefont {Mazur}, \citenamefont {Bordin}, \citenamefont {ten Haaf},
  \citenamefont {Wang}, \citenamefont {van Driel}, \citenamefont {Zatelli},
  \citenamefont {Li}, \citenamefont {Malinowski}, \citenamefont {Gazibegovic},
  \citenamefont {Badawy}, \citenamefont {Bakkers}, \citenamefont {Wimmer},\
  and\ \citenamefont {Kouwenhoven}}]{dvir_realization_2023}%
  \BibitemOpen
  \bibfield  {author} {\bibinfo {author} {\bibfnamefont {T.}~\bibnamefont
  {Dvir}}, \bibinfo {author} {\bibfnamefont {G.}~\bibnamefont {Wang}}, \bibinfo
  {author} {\bibfnamefont {N.}~\bibnamefont {van Loo}}, \bibinfo {author}
  {\bibfnamefont {C.-X.}\ \bibnamefont {Liu}}, \bibinfo {author} {\bibfnamefont
  {G.~P.}\ \bibnamefont {Mazur}}, \bibinfo {author} {\bibfnamefont
  {A.}~\bibnamefont {Bordin}}, \bibinfo {author} {\bibfnamefont {S.~L.~D.}\
  \bibnamefont {ten Haaf}}, \bibinfo {author} {\bibfnamefont {J.-Y.}\
  \bibnamefont {Wang}}, \bibinfo {author} {\bibfnamefont {D.}~\bibnamefont {van
  Driel}}, \bibinfo {author} {\bibfnamefont {F.}~\bibnamefont {Zatelli}},
  \bibinfo {author} {\bibfnamefont {X.}~\bibnamefont {Li}}, \bibinfo {author}
  {\bibfnamefont {F.~K.}\ \bibnamefont {Malinowski}}, \bibinfo {author}
  {\bibfnamefont {S.}~\bibnamefont {Gazibegovic}}, \bibinfo {author}
  {\bibfnamefont {G.}~\bibnamefont {Badawy}}, \bibinfo {author} {\bibfnamefont
  {E.~P. A.~M.}\ \bibnamefont {Bakkers}}, \bibinfo {author} {\bibfnamefont
  {M.}~\bibnamefont {Wimmer}},\ and\ \bibinfo {author} {\bibfnamefont {L.~P.}\
  \bibnamefont {Kouwenhoven}},\ }\bibfield  {title} {\bibinfo {title}
  {Realization of a minimal {Kitaev} chain in coupled quantum dots},\ }\href
  {https://doi.org/10.1038/s41586-022-05585-1} {\bibfield  {journal} {\bibinfo
  {journal} {Nature}\ }\textbf {\bibinfo {volume} {614}},\ \bibinfo {pages}
  {445} (\bibinfo {year} {2023})}\BibitemShut {NoStop}%
\bibitem [{\citenamefont {{ten Haaf}}\ \emph {et~al.}(2023)\citenamefont {{ten
  Haaf}}, \citenamefont {Wang}, \citenamefont {Bozkurt}, \citenamefont {Liu},
  \citenamefont {Kulesh}, \citenamefont {Kim}, \citenamefont {Xiao},
  \citenamefont {Thomas}, \citenamefont {Manfra}, \citenamefont {Dvir},
  \citenamefont {Wimmer},\ and\ \citenamefont
  {Goswami}}]{haaf_engineering_2023}%
  \BibitemOpen
  \bibfield  {author} {\bibinfo {author} {\bibfnamefont {S.~L.~D.}\
  \bibnamefont {{ten Haaf}}}, \bibinfo {author} {\bibfnamefont
  {Q.}~\bibnamefont {Wang}}, \bibinfo {author} {\bibfnamefont {A.~M.}\
  \bibnamefont {Bozkurt}}, \bibinfo {author} {\bibfnamefont {C.-X.}\
  \bibnamefont {Liu}}, \bibinfo {author} {\bibfnamefont {I.}~\bibnamefont
  {Kulesh}}, \bibinfo {author} {\bibfnamefont {P.}~\bibnamefont {Kim}},
  \bibinfo {author} {\bibfnamefont {D.}~\bibnamefont {Xiao}}, \bibinfo {author}
  {\bibfnamefont {C.}~\bibnamefont {Thomas}}, \bibinfo {author} {\bibfnamefont
  {M.~J.}\ \bibnamefont {Manfra}}, \bibinfo {author} {\bibfnamefont
  {T.}~\bibnamefont {Dvir}}, \bibinfo {author} {\bibfnamefont {M.}~\bibnamefont
  {Wimmer}},\ and\ \bibinfo {author} {\bibfnamefont {S.}~\bibnamefont
  {Goswami}},\ }\bibfield  {title} {\bibinfo {title} {Engineering {Majorana}
  bound states in coupled quantum dots in a two-dimensional electron gas},\
  }\href {http://arxiv.org/abs/2311.03208} {\bibfield  {journal} {\bibinfo
  {journal} {arXiv:2311.03208}} (\bibinfo {year} {2023})}\BibitemShut
  {NoStop}%
\bibitem [{\citenamefont {Zatelli}\ \emph {et~al.}(2023)\citenamefont
  {Zatelli}, \citenamefont {van Driel}, \citenamefont {Xu}, \citenamefont
  {Wang}, \citenamefont {Liu}, \citenamefont {Bordin}, \citenamefont {Roovers},
  \citenamefont {Mazur}, \citenamefont {van Loo}, \citenamefont {Wolff},
  \citenamefont {Bozkurt}, \citenamefont {Badawy}, \citenamefont {Gazibegovic},
  \citenamefont {Bakkers}, \citenamefont {Wimmer}, \citenamefont
  {Kouwenhoven},\ and\ \citenamefont {Dvir}}]{zatelli_robust_2023}%
  \BibitemOpen
  \bibfield  {author} {\bibinfo {author} {\bibfnamefont {F.}~\bibnamefont
  {Zatelli}}, \bibinfo {author} {\bibfnamefont {D.}~\bibnamefont {van Driel}},
  \bibinfo {author} {\bibfnamefont {D.}~\bibnamefont {Xu}}, \bibinfo {author}
  {\bibfnamefont {G.}~\bibnamefont {Wang}}, \bibinfo {author} {\bibfnamefont
  {C.-X.}\ \bibnamefont {Liu}}, \bibinfo {author} {\bibfnamefont
  {A.}~\bibnamefont {Bordin}}, \bibinfo {author} {\bibfnamefont
  {B.}~\bibnamefont {Roovers}}, \bibinfo {author} {\bibfnamefont {G.~P.}\
  \bibnamefont {Mazur}}, \bibinfo {author} {\bibfnamefont {N.}~\bibnamefont
  {van Loo}}, \bibinfo {author} {\bibfnamefont {J.~C.}\ \bibnamefont {Wolff}},
  \bibinfo {author} {\bibfnamefont {A.~M.}\ \bibnamefont {Bozkurt}}, \bibinfo
  {author} {\bibfnamefont {G.}~\bibnamefont {Badawy}}, \bibinfo {author}
  {\bibfnamefont {S.}~\bibnamefont {Gazibegovic}}, \bibinfo {author}
  {\bibfnamefont {E.~P. A.~M.}\ \bibnamefont {Bakkers}}, \bibinfo {author}
  {\bibfnamefont {M.}~\bibnamefont {Wimmer}}, \bibinfo {author} {\bibfnamefont
  {L.~P.}\ \bibnamefont {Kouwenhoven}},\ and\ \bibinfo {author} {\bibfnamefont
  {T.}~\bibnamefont {Dvir}},\ }\bibfield  {title} {\bibinfo {title} {Robust
  poor man's {Majorana} zero modes using {Yu}-{Shiba}-{Rusinov} states},\
  }\href {http://arxiv.org/abs/2311.03193} {\bibfield  {journal} {\bibinfo
  {journal} {arXiv:2311.03193}} (\bibinfo {year} {2023})}\BibitemShut
  {NoStop}%
\bibitem [{\citenamefont {Bordin}\ \emph
  {et~al.}(2024{\natexlab{a}})\citenamefont {Bordin}, \citenamefont {Li},
  \citenamefont {van Driel}, \citenamefont {Wolff}, \citenamefont {Wang},
  \citenamefont {ten Haaf}, \citenamefont {Wang}, \citenamefont {van Loo},
  \citenamefont {Kouwenhoven},\ and\ \citenamefont
  {Dvir}}]{bordin_crossed_2024}%
  \BibitemOpen
  \bibfield  {author} {\bibinfo {author} {\bibfnamefont {A.}~\bibnamefont
  {Bordin}}, \bibinfo {author} {\bibfnamefont {X.}~\bibnamefont {Li}}, \bibinfo
  {author} {\bibfnamefont {D.}~\bibnamefont {van Driel}}, \bibinfo {author}
  {\bibfnamefont {J.~C.}\ \bibnamefont {Wolff}}, \bibinfo {author}
  {\bibfnamefont {Q.}~\bibnamefont {Wang}}, \bibinfo {author} {\bibfnamefont
  {S.~L.~D.}\ \bibnamefont {ten Haaf}}, \bibinfo {author} {\bibfnamefont
  {G.}~\bibnamefont {Wang}}, \bibinfo {author} {\bibfnamefont {N.}~\bibnamefont
  {van Loo}}, \bibinfo {author} {\bibfnamefont {L.~P.}\ \bibnamefont
  {Kouwenhoven}},\ and\ \bibinfo {author} {\bibfnamefont {T.}~\bibnamefont
  {Dvir}},\ }\bibfield  {title} {\bibinfo {title} {Crossed {Andreev}
  {reflection} and {elastic} {cotunneling} in {three} {quantum} {dots}
  {coupled} by {superconductors}},\ }\href
  {https://doi.org/10.1103/PhysRevLett.132.056602} {\bibfield  {journal}
  {\bibinfo  {journal} {Phys. Rev. Lett.}\ }\textbf {\bibinfo {volume} {132}},\
  \bibinfo {pages} {056602} (\bibinfo {year} {2024}{\natexlab{a}})}\BibitemShut
  {NoStop}%
\bibitem [{\citenamefont {Bordin}\ \emph
  {et~al.}(2024{\natexlab{b}})\citenamefont {Bordin}, \citenamefont {Liu},
  \citenamefont {Dvir}, \citenamefont {Zatelli}, \citenamefont {Haaf},
  \citenamefont {van Driel}, \citenamefont {Wang}, \citenamefont {van Loo},
  \citenamefont {van Caekenberghe}, \citenamefont {Wolff}, \citenamefont
  {Zhang}, \citenamefont {Badawy}, \citenamefont {Gazibegovic}, \citenamefont
  {Bakkers}, \citenamefont {Wimmer}, \citenamefont {Kouwenhoven},\ and\
  \citenamefont {Mazur}}]{bordin_signatures_2024}%
  \BibitemOpen
  \bibfield  {author} {\bibinfo {author} {\bibfnamefont {A.}~\bibnamefont
  {Bordin}}, \bibinfo {author} {\bibfnamefont {C.-X.}\ \bibnamefont {Liu}},
  \bibinfo {author} {\bibfnamefont {T.}~\bibnamefont {Dvir}}, \bibinfo {author}
  {\bibfnamefont {F.}~\bibnamefont {Zatelli}}, \bibinfo {author} {\bibfnamefont
  {S.~L. D.~t.}\ \bibnamefont {Haaf}}, \bibinfo {author} {\bibfnamefont
  {D.}~\bibnamefont {van Driel}}, \bibinfo {author} {\bibfnamefont
  {G.}~\bibnamefont {Wang}}, \bibinfo {author} {\bibfnamefont {N.}~\bibnamefont
  {van Loo}}, \bibinfo {author} {\bibfnamefont {T.}~\bibnamefont {van
  Caekenberghe}}, \bibinfo {author} {\bibfnamefont {J.~C.}\ \bibnamefont
  {Wolff}}, \bibinfo {author} {\bibfnamefont {Y.}~\bibnamefont {Zhang}},
  \bibinfo {author} {\bibfnamefont {G.}~\bibnamefont {Badawy}}, \bibinfo
  {author} {\bibfnamefont {S.}~\bibnamefont {Gazibegovic}}, \bibinfo {author}
  {\bibfnamefont {E.~P. A.~M.}\ \bibnamefont {Bakkers}}, \bibinfo {author}
  {\bibfnamefont {M.}~\bibnamefont {Wimmer}}, \bibinfo {author} {\bibfnamefont
  {L.~P.}\ \bibnamefont {Kouwenhoven}},\ and\ \bibinfo {author} {\bibfnamefont
  {G.~P.}\ \bibnamefont {Mazur}},\ }\bibfield  {title} {\bibinfo {title}
  {Signatures of {Majorana} protection in a three-site {Kitaev} chain},\ }\href
  {http://arxiv.org/abs/2402.19382} {\bibfield  {journal} {\bibinfo  {journal}
  {arXiv:2402.19382}} (\bibinfo {year} {2024}{\natexlab{b}})}\BibitemShut
  {NoStop}%
\bibitem [{\citenamefont {Liu}\ \emph {et~al.}(2022)\citenamefont {Liu},
  \citenamefont {Wang}, \citenamefont {Dvir},\ and\ \citenamefont
  {Wimmer}}]{liu_tunable_2022}%
  \BibitemOpen
  \bibfield  {author} {\bibinfo {author} {\bibfnamefont {C.-X.}\ \bibnamefont
  {Liu}}, \bibinfo {author} {\bibfnamefont {G.}~\bibnamefont {Wang}}, \bibinfo
  {author} {\bibfnamefont {T.}~\bibnamefont {Dvir}},\ and\ \bibinfo {author}
  {\bibfnamefont {M.}~\bibnamefont {Wimmer}},\ }\bibfield  {title} {\bibinfo
  {title} {Tunable {superconducting} {coupling} of {quantum} {dots} via
  {Andreev} {bound} {states} in {semiconductor}-{superconductor} {nanowires}},\
  }\href {https://doi.org/10.1103/PhysRevLett.129.267701} {\bibfield  {journal}
  {\bibinfo  {journal} {Phys. Rev. Lett.}\ }\textbf {\bibinfo {volume} {129}},\
  \bibinfo {pages} {267701} (\bibinfo {year} {2022})}\BibitemShut {NoStop}%
\bibitem [{\citenamefont {Liu}\ \emph {et~al.}(2023{\natexlab{b}})\citenamefont
  {Liu}, \citenamefont {Bozkurt}, \citenamefont {Zatelli}, \citenamefont
  {Haaf}, \citenamefont {Dvir},\ and\ \citenamefont
  {Wimmer}}]{liu_enhancing_2023}%
  \BibitemOpen
  \bibfield  {author} {\bibinfo {author} {\bibfnamefont {C.-X.}\ \bibnamefont
  {Liu}}, \bibinfo {author} {\bibfnamefont {A.~M.}\ \bibnamefont {Bozkurt}},
  \bibinfo {author} {\bibfnamefont {F.}~\bibnamefont {Zatelli}}, \bibinfo
  {author} {\bibfnamefont {S.~L. D.~t.}\ \bibnamefont {Haaf}}, \bibinfo
  {author} {\bibfnamefont {T.}~\bibnamefont {Dvir}},\ and\ \bibinfo {author}
  {\bibfnamefont {M.}~\bibnamefont {Wimmer}},\ }\bibfield  {title} {\bibinfo
  {title} {Enhancing the excitation gap of a quantum-dot-based {Kitaev}
  chain},\ }\href {http://arxiv.org/abs/2310.09106} {\bibfield  {journal}
  {\bibinfo  {journal} {arXiv:2310.09106}} (\bibinfo {year}
  {2023}{\natexlab{b}})}\BibitemShut {NoStop}%
\bibitem [{\citenamefont {Thamm}\ and\ \citenamefont
  {Rosenow}(2023)}]{thamm_machine_2023}%
  \BibitemOpen
  \bibfield  {author} {\bibinfo {author} {\bibfnamefont {M.}~\bibnamefont
  {Thamm}}\ and\ \bibinfo {author} {\bibfnamefont {B.}~\bibnamefont
  {Rosenow}},\ }\bibfield  {title} {\bibinfo {title} {Machine {learning}
  {optimization} of {Majorana} {hybrid} {nanowires}},\ }\href
  {https://doi.org/10.1103/PhysRevLett.130.116202} {\bibfield  {journal}
  {\bibinfo  {journal} {Phys. Rev. Lett.}\ }\textbf {\bibinfo {volume} {130}},\
  \bibinfo {pages} {116202} (\bibinfo {year} {2023})}\BibitemShut {NoStop}%
\bibitem [{\citenamefont {Prada}\ \emph {et~al.}(2017)\citenamefont {Prada},
  \citenamefont {Aguado},\ and\ \citenamefont
  {San-Jose}}]{prada_measuring_2017}%
  \BibitemOpen
  \bibfield  {author} {\bibinfo {author} {\bibfnamefont {E.}~\bibnamefont
  {Prada}}, \bibinfo {author} {\bibfnamefont {R.}~\bibnamefont {Aguado}},\ and\
  \bibinfo {author} {\bibfnamefont {P.}~\bibnamefont {San-Jose}},\ }\bibfield
  {title} {\bibinfo {title} {Measuring {Majorana} nonlocality and spin
  structure with a quantum dot},\ }\href
  {https://doi.org/10.1103/PhysRevB.96.085418} {\bibfield  {journal} {\bibinfo
  {journal} {Phys. Rev. B}\ }\textbf {\bibinfo {volume} {96}},\ \bibinfo
  {pages} {085418} (\bibinfo {year} {2017})}\BibitemShut {NoStop}%
\bibitem [{\citenamefont {Clarke}(2017)}]{clarke_experimentally_2017}%
  \BibitemOpen
  \bibfield  {author} {\bibinfo {author} {\bibfnamefont {D.~J.}\ \bibnamefont
  {Clarke}},\ }\bibfield  {title} {\bibinfo {title} {Experimentally accessible
  topological quality factor for wires with zero energy modes},\ }\href
  {https://doi.org/10.1103/PhysRevB.96.201109} {\bibfield  {journal} {\bibinfo
  {journal} {Phys. Rev. B}\ }\textbf {\bibinfo {volume} {96}},\ \bibinfo
  {pages} {201109} (\bibinfo {year} {2017})}\BibitemShut {NoStop}%
\bibitem [{\citenamefont {Souto}\ \emph {et~al.}(2023)\citenamefont {Souto},
  \citenamefont {Tsintzis}, \citenamefont {Leijnse},\ and\ \citenamefont
  {Danon}}]{souto_probing_2023}%
  \BibitemOpen
  \bibfield  {author} {\bibinfo {author} {\bibfnamefont {R.~S.}\ \bibnamefont
  {Souto}}, \bibinfo {author} {\bibfnamefont {A.}~\bibnamefont {Tsintzis}},
  \bibinfo {author} {\bibfnamefont {M.}~\bibnamefont {Leijnse}},\ and\ \bibinfo
  {author} {\bibfnamefont {J.}~\bibnamefont {Danon}},\ }\bibfield  {title}
  {\bibinfo {title} {Probing {Majorana} localization in minimal {Kitaev} chains
  through a quantum dot},\ }\href
  {https://doi.org/10.1103/PhysRevResearch.5.043182} {\bibfield  {journal}
  {\bibinfo  {journal} {Phys. Rev. Research}\ }\textbf {\bibinfo {volume}
  {5}},\ \bibinfo {pages} {043182} (\bibinfo {year} {2023})}\BibitemShut
  {NoStop}%
\bibitem [{\citenamefont {Koch}\ \emph {et~al.}(2023)\citenamefont {Koch},
  \citenamefont {van Driel}, \citenamefont {Bordin}, \citenamefont {Lado},\
  and\ \citenamefont {Greplova}}]{koch_adversarial_2023}%
  \BibitemOpen
  \bibfield  {author} {\bibinfo {author} {\bibfnamefont {R.}~\bibnamefont
  {Koch}}, \bibinfo {author} {\bibfnamefont {D.}~\bibnamefont {van Driel}},
  \bibinfo {author} {\bibfnamefont {A.}~\bibnamefont {Bordin}}, \bibinfo
  {author} {\bibfnamefont {J.~L.}\ \bibnamefont {Lado}},\ and\ \bibinfo
  {author} {\bibfnamefont {E.}~\bibnamefont {Greplova}},\ }\bibfield  {title}
  {\bibinfo {title} {Adversarial {Hamiltonian} learning of quantum dots in a
  minimal {Kitaev} chain},\ }\href
  {https://doi.org/10.1103/PhysRevApplied.20.044081} {\bibfield  {journal}
  {\bibinfo  {journal} {Phys. Rev. Applied}\ }\textbf {\bibinfo {volume}
  {20}},\ \bibinfo {pages} {044081} (\bibinfo {year} {2023})}\BibitemShut
  {NoStop}%
\bibitem [{\citenamefont {Hansen}(2016)}]{hansen_cma_2023}%
  \BibitemOpen
  \bibfield  {author} {\bibinfo {author} {\bibfnamefont {N.}~\bibnamefont
  {Hansen}},\ }\bibfield  {title} {\bibinfo {title} {The {CMA} {evolution}
  {strategy}: {A} {tutorial}},\ }\href {http://arxiv.org/abs/1604.00772}
  {\bibfield  {journal} {\bibinfo  {journal} {arXiv:1604.00772}} (\bibinfo
  {year} {2016})}\BibitemShut {NoStop}%
\bibitem [{\citenamefont {Hansen}\ and\ \citenamefont
  {Ostermeier}(2001)}]{hansen_completely_2001}%
  \BibitemOpen
  \bibfield  {author} {\bibinfo {author} {\bibfnamefont {N.}~\bibnamefont
  {Hansen}}\ and\ \bibinfo {author} {\bibfnamefont {A.}~\bibnamefont
  {Ostermeier}},\ }\bibfield  {title} {\bibinfo {title} {Completely
  {derandomized} {self}-{adaptation} in {evolution} {strategies}},\ }\href
  {https://doi.org/10.1162/106365601750190398} {\bibfield  {journal} {\bibinfo
  {journal} {Evol. Comput.}\ }\textbf {\bibinfo {volume} {9}},\ \bibinfo
  {pages} {159} (\bibinfo {year} {2001})}\BibitemShut {NoStop}%
\bibitem [{\citenamefont {Hansen}\ \emph {et~al.}(2003)\citenamefont {Hansen},
  \citenamefont {Müller},\ and\ \citenamefont
  {Koumoutsakos}}]{hansen_reducing_2003}%
  \BibitemOpen
  \bibfield  {author} {\bibinfo {author} {\bibfnamefont {N.}~\bibnamefont
  {Hansen}}, \bibinfo {author} {\bibfnamefont {S.~D.}\ \bibnamefont
  {Müller}},\ and\ \bibinfo {author} {\bibfnamefont {P.}~\bibnamefont
  {Koumoutsakos}},\ }\bibfield  {title} {\bibinfo {title} {Reducing the {time}
  {complexity} of the {derandomized} {evolution} {strategy} with {covariance}
  {matrix} {adaptation} ({CMA}-{ES})},\ }\href
  {https://doi.org/10.1162/106365603321828970} {\bibfield  {journal} {\bibinfo
  {journal} {Evol. Comput.}\ }\textbf {\bibinfo {volume} {11}},\ \bibinfo
  {pages} {1} (\bibinfo {year} {2003})}\BibitemShut {NoStop}%
\bibitem [{Note1()}]{Note1}%
  \BibitemOpen
  \bibinfo {note} {Including small Coulomb interactions on the proximitized
  dots does not change the qualitative results~\cite
  {tsintzis_creating_2022}}\BibitemShut {NoStop}%
\bibitem [{Note2()}]{Note2}%
  \BibitemOpen
  \bibinfo {note} {We use the \protect \texttt {cma} Python library from
  \protect \href
  {https://pypi.org/project/cma/}{pypi.org/project/cma}}\BibitemShut {NoStop}%
\bibitem [{\citenamefont {Kouwenhoven}\ \emph {et~al.}(1997)\citenamefont
  {Kouwenhoven}, \citenamefont {Marcus}, \citenamefont {McEuen}, \citenamefont
  {Tarucha}, \citenamefont {Westervelt},\ and\ \citenamefont
  {Wingreen}}]{kouwenhoven_electron_1997}%
  \BibitemOpen
  \bibfield  {author} {\bibinfo {author} {\bibfnamefont {L.~P.}\ \bibnamefont
  {Kouwenhoven}}, \bibinfo {author} {\bibfnamefont {C.~M.}\ \bibnamefont
  {Marcus}}, \bibinfo {author} {\bibfnamefont {P.~L.}\ \bibnamefont {McEuen}},
  \bibinfo {author} {\bibfnamefont {S.}~\bibnamefont {Tarucha}}, \bibinfo
  {author} {\bibfnamefont {R.~M.}\ \bibnamefont {Westervelt}},\ and\ \bibinfo
  {author} {\bibfnamefont {N.~S.}\ \bibnamefont {Wingreen}},\ }\bibfield
  {title} {\bibinfo {title} {Electron {Transport} in {Quantum} {Dots}},\ }in\
  \href {https://doi.org/10.1007/978-94-015-8839-3_4} {\emph {\bibinfo
  {booktitle} {Mesoscopic {Electron} {Transport}}}},\ \bibinfo {editor} {edited
  by\ \bibinfo {editor} {\bibfnamefont {L.~L.}\ \bibnamefont {Sohn}}, \bibinfo
  {editor} {\bibfnamefont {L.~P.}\ \bibnamefont {Kouwenhoven}},\ and\ \bibinfo
  {editor} {\bibfnamefont {G.}~\bibnamefont {Schön}}}\ (\bibinfo  {publisher}
  {Springer Netherlands},\ \bibinfo {address} {Dordrecht},\ \bibinfo {year}
  {1997})\ pp.\ \bibinfo {pages} {105--214}\BibitemShut {NoStop}%
\bibitem [{\citenamefont {Sedlmayr}\ and\ \citenamefont
  {Bena}(2015)}]{sedlmayr_visualizing_2015}%
  \BibitemOpen
  \bibfield  {author} {\bibinfo {author} {\bibfnamefont {N.}~\bibnamefont
  {Sedlmayr}}\ and\ \bibinfo {author} {\bibfnamefont {C.}~\bibnamefont
  {Bena}},\ }\bibfield  {title} {\bibinfo {title} {Visualizing {Majorana} bound
  states in one and two dimensions using the generalized {Majorana}
  polarization},\ }\href {https://doi.org/10.1103/PhysRevB.92.115115}
  {\bibfield  {journal} {\bibinfo  {journal} {Phys. Rev. B}\ }\textbf {\bibinfo
  {volume} {92}},\ \bibinfo {pages} {115115} (\bibinfo {year}
  {2015})}\BibitemShut {NoStop}%
\bibitem [{\citenamefont {Sticlet}\ \emph {et~al.}(2012)\citenamefont
  {Sticlet}, \citenamefont {Bena},\ and\ \citenamefont
  {Simon}}]{sticlet_spin_2012}%
  \BibitemOpen
  \bibfield  {author} {\bibinfo {author} {\bibfnamefont {D.}~\bibnamefont
  {Sticlet}}, \bibinfo {author} {\bibfnamefont {C.}~\bibnamefont {Bena}},\ and\
  \bibinfo {author} {\bibfnamefont {P.}~\bibnamefont {Simon}},\ }\bibfield
  {title} {\bibinfo {title} {Spin and {Majorana} {polarization} in
  {topological} {superconducting} {wires}},\ }\href
  {https://doi.org/10.1103/PhysRevLett.108.096802} {\bibfield  {journal}
  {\bibinfo  {journal} {Phys. Rev. Lett.}\ }\textbf {\bibinfo {volume} {108}},\
  \bibinfo {pages} {096802} (\bibinfo {year} {2012})}\BibitemShut {NoStop}%
\bibitem [{\citenamefont {Aksenov}\ \emph {et~al.}(2020)\citenamefont
  {Aksenov}, \citenamefont {Zlotnikov},\ and\ \citenamefont
  {Shustin}}]{aksenov_strong_2020}%
  \BibitemOpen
  \bibfield  {author} {\bibinfo {author} {\bibfnamefont {S.~V.}\ \bibnamefont
  {Aksenov}}, \bibinfo {author} {\bibfnamefont {A.~O.}\ \bibnamefont
  {Zlotnikov}},\ and\ \bibinfo {author} {\bibfnamefont {M.~S.}\ \bibnamefont
  {Shustin}},\ }\bibfield  {title} {\bibinfo {title} {Strong {Coulomb}
  interactions in the problem of {Majorana} modes in a wire of the nontrivial
  topological class {BDI}},\ }\href
  {https://doi.org/10.1103/PhysRevB.101.125431} {\bibfield  {journal} {\bibinfo
   {journal} {Phys. Rev. B}\ }\textbf {\bibinfo {volume} {101}},\ \bibinfo
  {pages} {125431} (\bibinfo {year} {2020})}\BibitemShut {NoStop}%
\bibitem [{Note3()}]{Note3}%
  \BibitemOpen
  \bibinfo {note} {Although such residual $\delta E_{eo}$ would still translate
  into a relatively short upper-bound time scale for ``Majorana manipulation,''
  it lies well within the resolution of an actual tunnelling-spectroscopy
  experiment, likely making it impossible to resolve such small energy
  splitting anyway. Furthermore, the automated tuning found here could serve as
  the starting point for a finer search using other methods.}\BibitemShut
  {Stop}%
\bibitem [{\citenamefont {Själander}\ \emph {et~al.}(2022)\citenamefont
  {Själander}, \citenamefont {Jahre}, \citenamefont {Tufte},\ and\
  \citenamefont {Reissmann}}]{sjalander_epic_2022}%
  \BibitemOpen
  \bibfield  {author} {\bibinfo {author} {\bibfnamefont {M.}~\bibnamefont
  {Själander}}, \bibinfo {author} {\bibfnamefont {M.}~\bibnamefont {Jahre}},
  \bibinfo {author} {\bibfnamefont {G.}~\bibnamefont {Tufte}},\ and\ \bibinfo
  {author} {\bibfnamefont {N.}~\bibnamefont {Reissmann}},\ }\bibfield  {title}
  {\bibinfo {title} {{EPIC}: {An} energy-efficient, high-performance gpgpu
  computing research infrastructure},\ }\href {http://arxiv.org/abs/1912.05848}
  {\bibfield  {journal} {\bibinfo  {journal} {arXiv:1912.05848}} (\bibinfo
  {year} {2022})}\BibitemShut {NoStop}%
\end{thebibliography}
\end{document}